# Direct Observation of Chirality-Induced Spin Selectivity in Electron Donor-Acceptor Molecules

Hannah J. Eckvahl,[1†] Nikolai A. Tcyrulnikov,[1†] Alessandro Chiesa,[2†] Jillian M. Bradley,[1] Ryan M. Young,[1] Stefano Carretta,[2*] Matthew D. Krzyaniak,[1*] and Michael R. Wasielewski[1*]

[1]Department of Chemistry, Center for Molecular Quantum Transduction, and Paula M. Trienens Institute for Sustainability and Energy, Northwestern University, Evanston, IL 60208-3113 USA.
[2]Università di Parma, Dipartimento di Scienze Matematiche, Fisiche e Informatiche, Parma, I-43124, Italy.

*Corresponding authors. Email: m-wasielewski@northwestern.edu; mdkrzyaniak@northwestern.edu; stefano.carretta@unipr.it

[†]These authors contributed equally to this work

**One-Sentence Summary:** The spin dynamics of photogenerated radical pairs are strongly influenced by molecular chirality.

**Abstract:**
The role of chirality in determining the spin dynamics of photoinduced electron transfer in donor-acceptor molecules remains an open question. Although chirality-induced spin selectivity (CISS) has been demonstrated in molecules bound to substrates, experimental information about whether this process influences spin dynamics in the molecules themselves is lacking. Here we use time-resolved electron paramagnetic resonance spectroscopy to show that CISS strongly influences the spin dynamics of isolated covalent donor-chiral bridge-acceptor (D-Bχ-A) molecules in which selective photoexcitation of D is followed by two rapid, sequential electron transfer events to yield $D^{\bullet+}$-Bχ-$A^{\bullet-}$. Exploiting this phenomenon affords the possibility of using chiral molecular building blocks to control electron spin states in quantum information applications.

Molecules offer a wide variety of quantum properties that could potentially be exploited in qubit architectures for quantum information science (QIS).(*1, 2*) Moreover, molecules afford the ability to tailor these properties as applications dictate, while controlling structure with atomic precision. One such property of growing interest is molecular chirality, which plays an essential role in many chemical reactions and nearly all biological processes. Naaman, Waldeck, and coworkers presented evidence of the relationship between molecular chirality and electron spin (*3, 4*) when they observed that electrons photoemitted from a gold surface coated with a thin film of DNA have a preferred spin state, a phenomenon now known as chirality-induced spin selectivity (CISS).(*5*) Subsequent experiments using molecules bound to metallic, semiconductor, or magnetic substrates have confirmed a connection between electron motion and spin projection along the chiral axis, which is selected to be parallel or antiparallel to the motion, depending on the handedness of the chiral molecule.(*5-9*) The spin selectivity of the effect can be very high, even at room temperature, and its theoretical foundations are still being explored.(*10-17*) However, a key problem hindering a fundamental understanding of CISS is that it is difficult to separate the role of the substrate from that of the chiral molecule.

Hence, it is crucial to investigate how CISS affects electron spin dynamics in molecules undergoing electron transfer that are not bound to substrates. Achieving this understanding would make it possible to design chiral molecular building blocks to manipulate electron spin states, which has potential for QIS applications. In particular, the occurrence of CISS at the molecular level has been proposed as an enabling technology for quantum applications, e.g. solving key issues like single-spin readout and high-temperature spin qubit initialization.(*6*)

Here we show direct evidence of CISS in isolated covalent donor-chiral bridge-acceptor (D-B$\chi$-A) molecules in which selective photoexcitation of D to its lowest excited singlet state ($^{1*}$D) is followed by two rapid, sequential electron transfer events: $^{1*}$D-B$\chi$-A $\rightarrow$ D$^{\bullet+}$-B$\chi^{\bullet-}$-A $\rightarrow$ D$^{\bullet+}$-B$\chi$-A$^{\bullet-}$ (Fig. 1A). If formation of D$^{\bullet+}$-B$\chi$-A$^{\bullet-}$ occurs in $\lesssim$1 ns, the resulting entangled electron spin pair is prepared initially in a nearly pure singlet state, $^{1}$(D$^{\bullet+}$-B$\chi$-A$^{\bullet-}$). These states are commonly referred to as spin-correlated radical pairs (SCRPs) and have been studied in systems ranging from photosynthetic proteins(*18-21*) and related model systems(*22-25*) to DNA hairpins.(*26-30*) However, in all these cases, no consideration was given to the possible influence of chirality on the SCRP spin dynamics.

To demonstrate the occurrence of CISS, we have synthesized pairs of covalent D-B$\chi$-A enantiomers, (*R*)-**1-*h*₉** (**-*d*₉**) and (*S*)-**1-*h*₉** (**-*d*₉**), where D is either non-deuterated (**-*h*₉**) or fully deuterated (**-*d*₉**) *peri*-xanthenoxanthene,(*31*) B$\chi$ is a pair of naphthalene-1,8-dicarboximides that are linked at their 4-positions to form an enantiomeric pair of axially chiral dimers (*R*)-NMI$_2$ and (*S*)-NMI$_2$(*32*) and A is naphthalene-1,8:4,5-bis(dicarboximide) (NDI) (see Supplementary Materials, figs. S1-S2). The structures of (*R*)-**1-*h*₉** (**-*d*₉**) and (*S*)-**1-*h*₉** (**-*d*₉**) and the corresponding achiral reference molecules **2-*h*₉** (**-*d*₉**) are shown in Fig. 1B. The enantiomers were separated using HPLC with a chiral stationary phase (fig. S3) and their circular dichroism spectra are given in fig. S4. We have characterized the charge separation and recombination dynamics of these molecules using transient optical absorption (TA) spectroscopy and the CISS effect on their spin dynamics using time-resolved EPR spectroscopy using either continuous (TREPR) or pulsed microwave radiation (pulse-EPR).

We find that CISS yields characteristic features in the TREPR spectra of the photogenerated PXX$^{\bullet+}$- NMI$_2$-NDI$^{\bullet-}$ SCRP in (*R*)-**1-*h*₉** (**-*d*₉**) and (*S*)-**1-*h*₉** (**-*d*₉**), which are absent in achiral **2-*h*₉** (**-*d*₉**), when the direction of electron transfer is oriented orthogonal to the applied static magnetic field direction in agreement with simulations. Conversely, the corresponding spectra of PXX$^{•+}$-NMI$_2$-NDI$^{•-}$ are practically identical when the field is parallel to the electron transfer direction.

**Time-resolved EPR Spectroscopy**

Samples of (*R*)-**1-*h*₉** (**-*d*₉**), (*S*)-**1-*h*₉** (**-*d*₉**), and **2-*h*₉** (**-*d*₉**) were each prepared in the nematic liquid crystal 4-cyano-4'-(*n*-pentyl)biphenyl (5CB), which was aligned in a magnetic field at 295 K, then rapidly frozen to 85 K, which aligns the long axes of these molecules along the magnetic field. The orientation of the molecules aligned in frozen 5CB can then be rotated relative to the applied magnetic field direction. Since solid 5CB is an optically scattering medium, to assess the photo-driven charge separation dynamics of these molecules at low temperature, we employed both femtosecond and nanosecond transient optical absorption spectroscopy substituting glassy butyronitrile for 5CB at 105 K. Transient absorption spectra and kinetics are given in figs. S5 and S6. The data show that in each case ultrafast two-step charge separation occurs in $\lesssim$ 0.2 ns to give PXX$^{•+}$-NDI$^{•-}$, which recombines to its ground state in $\tau$ = 46-66 μs, providing ample time for TREPR measurements. The presence of a ~0.35 T static magnetic field in the TREPR experiments does not affect the ultrafast electron transfer reactions because the Zeeman interaction (~0.3 cm$^{-1}$) at that field strength is much less than the adiabatic energy gaps (~20-80 cm$^{-1}$) for these reactions (see table S1 and SM for details).

Pulse-EPR techniques were used to assess the quality of the alignment of (*R*)-**1-*h*₉** (**-*d*₉**), (*S*)-**1-*h*₉** (**-*d*₉**), and **2-*h*₉** (**-*d*₉**) 5CB by measuring the isotropic exchange ($J$) and dipolar ($D$) spin-spin interactions for their photogenerated SCRPs, where $D(\theta) = d(1 - 3\cos^2\theta)$ and $d = 52.04\ \text{MHz} \cdot \text{nm}^3/r_{DA}^3$ in the point dipole approximation, which gives detailed distance and orientation information as defined by the Hamiltonian in eq. S3. If photogeneration of the SCRP is followed by a Hahn echo microwave pulse sequence: π/2 pulse - delay τ - π pulse - delay τ - spin echo, and the time delay τ is scanned, coherent oscillations between the eigenstates of the SCRP Hamiltonian $|\Phi_A\rangle$ and $|\Phi_B\rangle$ (see below) that are related to both $J$ and $D(\theta)$ modulate the spin echo amplitude.(*21, 33-37*) When this experiment is performed on spin-coherent SCRPs, the echo appears out-of-phase, *i.e.*, in the detection channel in quadrature to the one in which it is expected, and is therefore termed out-of-phase electron spin echo envelope modulation (OOP-ESEEM).(*21, 33-37*) For large SCRP distances, $r_{DA}$, $J$ can be neglected and the OOP-ESEEM oscillation frequency is approximately 2$d$ when $\theta$ = 0° and $d$ when $\theta$ = 90°. Thus, OOP-ESEEM can be used to measure SCRP distances for a given angle of the dipolar axis relative to the magnetic field.(*28, 36-38*) The dipolar axis in SCRPs connects the centroids of the spin distributions of the two radicals. Figures S7 and S8 show the OOP-ESEEM data for (*R*)-**1-*h*₉** (**-*d*₉**), (*S*)-**1-*h*₉** (**-*d*₉**), and **2-*h*₉** (**-*d*₉**) assuming that their dipolar axes are aligned parallel or perpendicular to the magnetic field. Fitting the OOP-ESEEM data shows that the measured PXX-*h*₉$^{•+}$-NDI$^{•-}$ distances of (*R*)-**1-*h*₉**, (*S*)-**1-*h*₉**, and **2-*h*₉** are 2.48 ± 0.01, 2.48 ± 0.01, and 2.28 ± 0.01 nm, respectively, while the corresponding PXX-*d*₉$^{•+}$-NDI$^{•-}$ distances of (*R*)-**1-*d*₉**, (*S*)-**1-*d*₉**, and **2-*d*₉** are 2.53 ± 0.01, 2.51 ± 0.01, and 2.29 ± 0.02 nm, respectively (table S2). These experimental distances are consistent with the center-to-center distances between PXX and NDI determined from density functional theory calculations on (*R*)-**1-*h*₉**, (*S*)-**1-*h*₉**, and **2-*h*₉**, where $r_{DA}$ = 2.59, 2.60, and 2.40 nm, respectively (fig. S9 and tables S3-S5). The agreement between the experimental and calculated distances shows that the D$^{•+}$-Bχ-A$^{•-}$ SCRPs are well-aligned along the magnetic field direction in frozen 5CB.

The TREPR spectra of aligned (*R*)-**1-*h*₉**, (*S*)-**1-*h*₉**, and **2-*h*₉** were obtained by photoexciting the samples with a 450 nm, 7 ns laser pulse and monitoring the magnetization with continuous microwaves using direct detection (see SM). The spectra obtained 100 ns after the laser pulse ($T_{DAF}$) are shown in Fig. 2. When the long axes of these molecules are aligned parallel to the magnetic field direction ($\theta$ = 0°), both enantiomers as well as the achiral reference molecule give the same spectra (Figs. 2A and 2C). Rotating the samples so that the long axes of ($R$)-**1-*h*9**, ($S$)-**1-*h*9**, and **2-*h*9** are aligned perpendicular to the magnetic field direction ($\theta$ = 90°), results in the appearance of outer wings in the spectra of chiral ($R$)-**1-*h*9** and ($S$)-**1-*h*9** (Fig. 2B). No such enhancement is observed for achiral **2-*h*9**. As explained below, we posit that these new features result from the contribution of CISS to the formation of the SCRPs in ($R$)-**1-*h*9** and ($S$)-**1-*h*9**. Deuteration of $PXX^{\bullet+}$ narrows the overall linewidth of ($R$)-**1-*d*9**, ($S$)-**1-*d*9**, and **2-*d*9**, while retaining the same orientation dependence of the signal (Figs. 2C and 2D).

**Effect of CISS on Radical Pair Spin Dynamics**

In the molecules described here, the $D^{\bullet+}$-$A^{\bullet-}$ distances are ≳23 Å, so that the spin-spin interactions $J$ and $D$ are small relative to the ~0.35 T applied magnetic field. Thus, the Zeeman term is by far the leading term in the SCRP spin Hamiltonian (eq. S3), so that the SCRP wavefunctions $|S\rangle = \frac{1}{\sqrt{2}}(|\uparrow\downarrow\rangle - |\downarrow\uparrow\rangle)$ and $|T_0\rangle = \frac{1}{\sqrt{2}}(|\uparrow\downarrow\rangle + |\downarrow\uparrow\rangle)$, which are magnetic field invariant, remain close in energy, while $|T_{+1}\rangle = |\uparrow\uparrow\rangle$ and $|T_{-1}\rangle = |\downarrow\downarrow\rangle$ are well separated in energy from both $|S\rangle$ and $|T_0\rangle$. In particular, both $|T_{+1}\rangle$ and $|T_{-1}\rangle$ are eigenstates of the spin Hamiltonian, while $|S\rangle$ and $|T_0\rangle$ are not eigenstates because of the different electronic $g$-factors and hyperfine fields of the two spins. Coherent mixing of $|S\rangle$ and $|T_0\rangle$ yields $|\Phi_A\rangle = \cos\phi\,|S\rangle + \sin\phi\,|T_0\rangle$ and $|\Phi_B\rangle = -\sin\phi\,|S\rangle + \cos\phi\,|T_0\rangle$ (Fig. 3A), which are eigenstates of the spin Hamiltonian, where the angle $\phi$ in the mixing coefficients is derived from the magnetic parameters of the SCRP (see SM).(*39-41*)

In the ultrafast electron transfer regime observed here, the initial spin state for an achiral SCRP is the entangled singlet $|S\rangle$ state yielding populations only on $|\Phi_A\rangle$ and $|\Phi_B\rangle$. Therefore, four allowed transitions occur between them and the initially unpopulated $|T_{+1}\rangle$ and $|T_{-1}\rangle$ states, giving rise to a spin-polarized (out of equilibrium) EPR spectrum. When $\theta$ = 0°, this results in a typical (*e*, *a*, *e*, *a*) spin polarization pattern (low to high field, where *a* = enhanced absorption, and *e* = emission) because $D(\theta) < 0$. (*39-41*) Conversely, when $\theta$ = 90°, the pattern is reversed (*a*, *e*, *a*, *e*) because $D(\theta) > 0$. Since the $g$-tensors of $PXX^{\bullet+}$ and $NDI^{\bullet-}$ are very similar, i.e. [2.0045, 2.0045, 2.0031](*42*) and [2.0047 2.0047 2.0027],(*43*) respectively, the expected SCRP polarization patterns, (*e*, *a*, *e*, *a*) or (*a*, *e*, *a*, *e*), are reduced to broadened (*e*, *a*) or (*a*, *e*) patterns, as observed experimentally for the achiral reference molecules **2-*h*9** and **2-*d*9** (Fig. 2, blue traces). Our OOP-ESEEM results show that the dipolar axis of each SCRP is well aligned with the long axis of each molecule so that the dipolar axis and the chirality axis of ($R$)-**1-*h*9** (**-*d*9**) and ($S$)-**1-*h*9** (**-*d*9**) are nearly parallel. The angle $\theta$ between this axis and the applied magnetic field ($B_0$) direction is depicted in Fig. 3B for the parallel and in Fig. 3E for perpendicular orientations. CISS mixes triplet character into the initial singlet SCRP, thus the initial populations of $|\Phi_A\rangle$, $|\Phi_B\rangle$, $|T_{+1}\rangle$ and $|T_{-1}\rangle$ and the corresponding transition intensities are predicted to change as well.(*44-47*) If CISS is the sole contribution to the spin dynamics and $\theta$ = 0° (Fig. 3D), the state following electron transfer would be $|\uparrow\downarrow\rangle$ or $|\downarrow\uparrow\rangle$, depending on the chirality of the enantiomer and whether $B_0$ is parallel or antiparallel to the electron motion. Given that the typical alignment of linear D-B-A molecules within nematic liquid crystals is not unidirectional, $B_0$ has equal probability of being parallel or antiparallel to the electron motion and hence, if coherences are lost, the initial state is an equal mixture of $|\uparrow\downarrow\rangle$ and $|\downarrow\uparrow\rangle$, and is thus equivalent to having a pure initial $|S\rangle$ state. This situation is shown schematically in Fig. 3C, where the blue and red traces depict the idealized TREPR spectra expected when CISS contributes 0 and 100%, respectively. Indeed, the observed spectra of both enantiomers as well as the achiral reference molecule are practically identical for $\theta = 0°$ (Figs. 2A and 2C).

In contrast, when the chirality axis is orthogonal to $B_0$ (Fig. 3E) the initial state is very different in the presence or absence of CISS. In particular, the CISS contribution initially populates $|T_{+1}\rangle$ and $|T_{-1}\rangle$ (see eq. S12). Therefore, if the SCRP spin state has a 100% CISS contribution, the TREPR spectra have a nearly opposite intensity pattern with respect to the case in which CISS does not contribute. This is illustrated in Fig. 3F where the blue and red lines in the idealized TREPR spectra correspond to the intensity for the pure $|S\rangle$ ($I_S$) and pure CISS ($I_{CISS}$) initial conditions, respectively.

Starting from recent theoretical models(*44-47*) describing the influence of CISS on SCRP spin dynamics in cases for which the CISS contribution will not be 100%, the initial state will be a superposition or a mixture of $|S\rangle$ and $|T_0\rangle$ along the chiral axis direction, making the detection of CISS less obvious. In fact, the spectral line intensities in this case are the weighted sum of $I_S$ and $I_{CISS}$, which occur at the same resonance fields and tend to cancel out (see details in the SM). The key to unraveling CISS and pure singlet contributions to the SCRP spin state in the molecules studied here is the observation of a larger EPR linewidth that occurs when CISS contributes. Indeed, the sum of the two contributions (Fig. 3F, black trace) yields a signal that displays lateral wings of opposite sign and central features that are narrower than those produced in the absence of CISS, exactly as observed experimentally in Figs. 2B and 2D. These features are unambiguous signatures of CISS, because they cannot be produced starting from an initial $|S\rangle$ state, where the polarization pattern is fixed to ($a$, $e$) for $\theta = 90°$ by the sign of $D(\theta)$.(*39-41*)

The larger linewidth obtained for the CISS initial state arises from the very different dependence of $I_S$ and $I_{CISS}$ on the degree of coherent mixing in the eigenstates. Indeed, $|I_{CISS}|$ increases with increasing entanglement ($\phi \rightarrow 0$) while $|I_S|$ decreases (fig. S10). Exploring the variation of the intensity by varying the composition of the $|\Phi_A\rangle$ and $|\Phi_B\rangle$ eigenstates is made possible by the presence of several nuclear spins and by distributed magnetic parameters, e.g. dipolar couplings, often termed strain. Therefore, moving from the center of each transition, i.e. the center of the distributions of the magnetic parameters and hyperfine fields, to the tail of the lineshape corresponds to changing the composition of the eigenstates, producing different linewidths for different initial states. If entanglement in the eigenstates is larger in the tails of the spectrum, the CISS contributions result in magnetic field dependent broadening, giving rise to lateral contributions to the lineshape of opposite sign with respect to the central features (Fig. 3F, black trace).

To confirm this interpretation, we consider the spectra of partially deuterated (*R*)-**1-*d*₉** and (*S*)-**1-*d*₉**. By strongly diminishing the hyperfine couplings on one of the two radicals, we change the distribution of the eigenstate composition and probe its effect on the lineshape. The measured spectra for $\theta = 0°$ and $\theta = 90°$ are shown in Figs. 2C and 2D, respectively. While no qualitative effect is visible in the parallel direction as expected, the lateral wings are significantly reduced in the perpendicular orientation. These spectra were simulated using a minimal SCRP model with either one spin-½ nucleus (hydrogen atom) on both $NDI^{\bullet-}$ and $PXX^{+}$ or only on $NDI^{\bullet-}$, the latter of which is the partially deuterated case. For reasonable values of the hyperfine couplings, the simulations shown in Fig. 4 reproduce the experimental behavior.

The intensities of the lateral wings are correctly reproduced by combining $I_S$ and $I_{CISS}$ with weights of 41% and 59%, respectively (Fig. 4). While a 59% CISS contribution is remarkable, it must be stressed that this is a minimal model in which the effect of the nuclei is accounted for only qualitatively and a full spectral simulation with all nuclear spins in the fully protonated molecules is very demanding. However, we have been able to perform the simulation for the deuterated case, which includes all four $^{1}$H and two $^{14}$N nuclei coupled to the electron spin in NDI$^{•-}$ and effects of dipolar strain. In this case, the experimental behavior is very well reproduced with a 47% CISS contribution, which is still considerable.

Further evidence for the validity of this interpretation is obtained by investigating the time dependence of the TREPR spectra, which reflect the time evolution of the D$^{•+}$-Bχ-A$^{•-}$ spin states under the combined effect of coherent and incoherent terms as described by the stochastic Liouville equation and the presence of the microwave field (see SM). Indeed, figs. S11 and S12 show that the dependence of the observed intensity of the wings of the spectra are similar to that of the main peaks, in agreement with our numerical simulations.

Interestingly, the CISS contribution to the spin dynamics of (*R*)-**1-*h*$_9$ (-*d*$_9$)** and (*S*)-**1-*h*$_9$ (-*d*$_9$)** is similar to the ~50% spin polarization recently reported for an axially chiral binaphthalene derivative covalently linked to a gold film deposited on nickel.(*48*) While this single comparison suggests that the observed CISS effect for the binaphthalene attached to the gold surface may be largely due to the chiral molecule, additional comparative work is needed on a variety of systems to warrant such a conclusion.

## Conclusions

We have found direct evidence of the CISS effect on the spin dynamics of photogenerated radical ion pairs in molecular electron donor-acceptor molecules. The observation of CISS in these systems affords possibilities both for increasing our understanding of this important phenomenon and for its possible applications. Importantly, these results show that the substrates/electrodes with their possibly large spin-orbit couplings are not needed for CISS to occur and that TREPR spectroscopy can directly access the spin dynamics resulting from CISS. This provides key information to guide theoretical investigations and makes possible many new targeted experimental studies. In addition, observing CISS at the molecular level is the first step required to transform this fundamental phenomenon into an enabling technology for quantum applications.

**Acknowledgments:**

**Funding:**
This work was supported by the National Science Foundation under award no. CHE-2154627 (M.R.W., synthesis, transient optical and EPR measurements).
Research supported as part of the Center for Molecular Quantum Transduction, an Energy Frontier Research Center funded by the U.S. Department of Energy (DOE), Office of Science, Basic Energy Sciences (BES), under award DE-SC0021314 (M.D.K., EPR data analysis).
This work was supported by the ERC-Synergy project CASTLE (project no. 101071533) funded by the Horizon Europe Program and the Fondazione Cariparma (S.C., calculations).
$^{1}$H nuclear magnetic resonance (NMR) spectroscopy, and mass spectrometry are conducted in IMSERC facilities at Northwestern University, which have received support from the Soft and Hybrid Nanotechnology Experimental (SHyNE) Resource (NSF ECCS-2025633), NSF CHE-1048773, Northwestern University, the State of Illinois, and the International Institute for Nanotechnology (IIN).

**Author contributions:**
Conceptualization: MRW
Methodology: HJE, NAT, JMB, MDK, AC, SC, RMY, MRW
Investigation: HJE, NAT, JMB, MDK, AC, SC, RMY, MRW
Visualization: HJE, NAT, AC, RMY, MRW
Funding acquisition: MRW, SC
Project administration: MRW
Supervision: MRW, MDK, SC
Writing – original draft: MRW, MDK, HJE, NAT, SC, AC
Writing – review & editing: MRW, MDK, HJE, NAT, SC, AC

**Competing interests:**
Authors declare that they have no competing interests.

**Data and materials availability:**
All data are available in the main text, the supplementary materials, and in Dryad(*49*).

**Supplementary Materials**
Materials and Methods
Supplementary Text
Figs. S1 to S12
Tables S1 to S7
References (*50-60*)

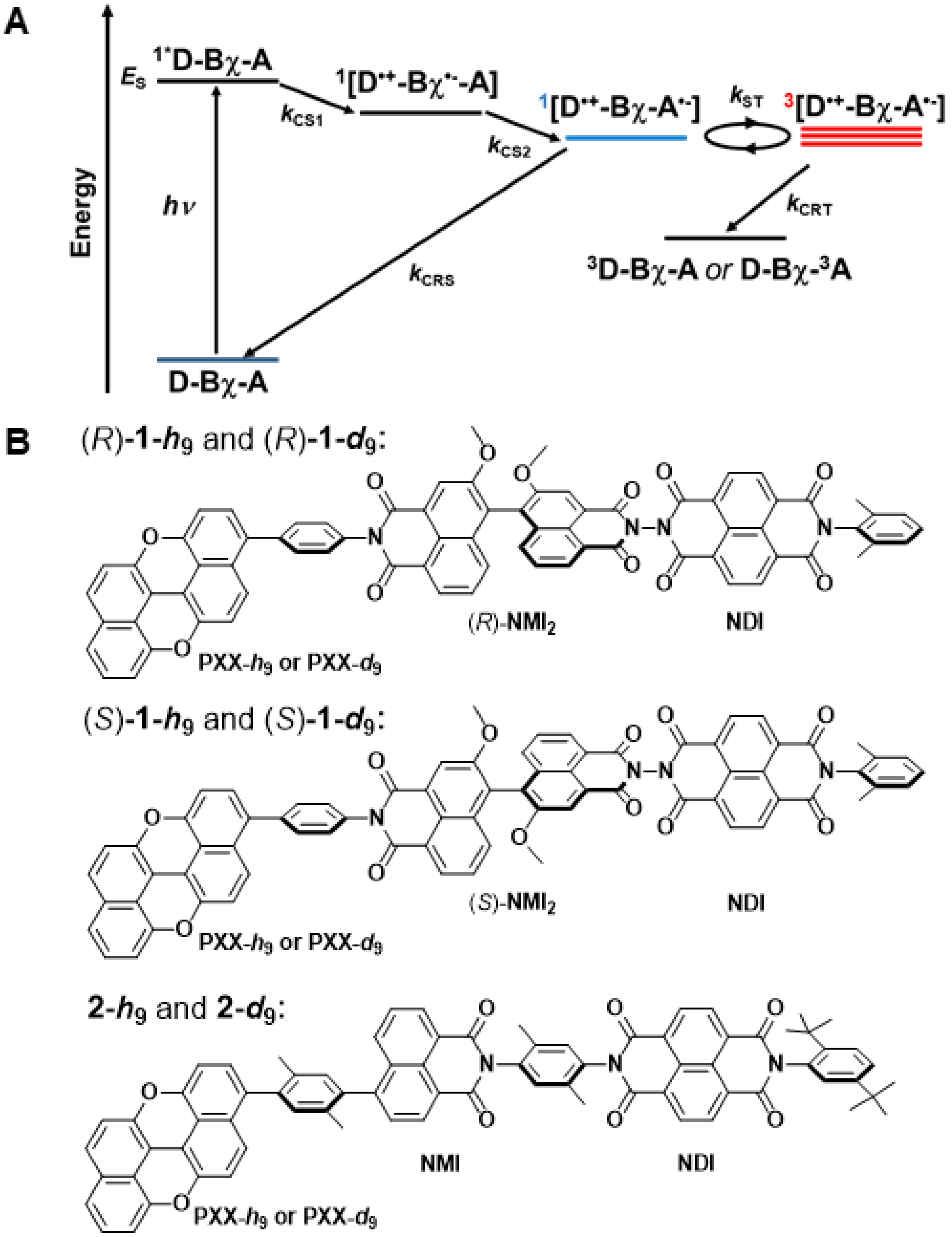

**Fig. 1. Electron transfer pathways and molecular structures.** (**A**) Electron transfer and intersystem crossing pathways in a D-Bχ-A system with no applied magnetic field, where $k_{CS1}$ and $k_{CS2}$ are the charge separation rate constants, $k_{ST}$ is singlet-triplet mixing rate constant, and $k_{CRS}$ and $k_{CRT}$ are the charge recombination rates via the singlet and triplet channels, respectively. (**B**) Structures of chiral (*R*)-**1** and **(***S***)**-**1** and achiral **2**. The steric constraints imposed by linking the two NMI groups in (*R*)-**1** and **(***S***)**-**1** result in stable enantiomers having axial chirality.

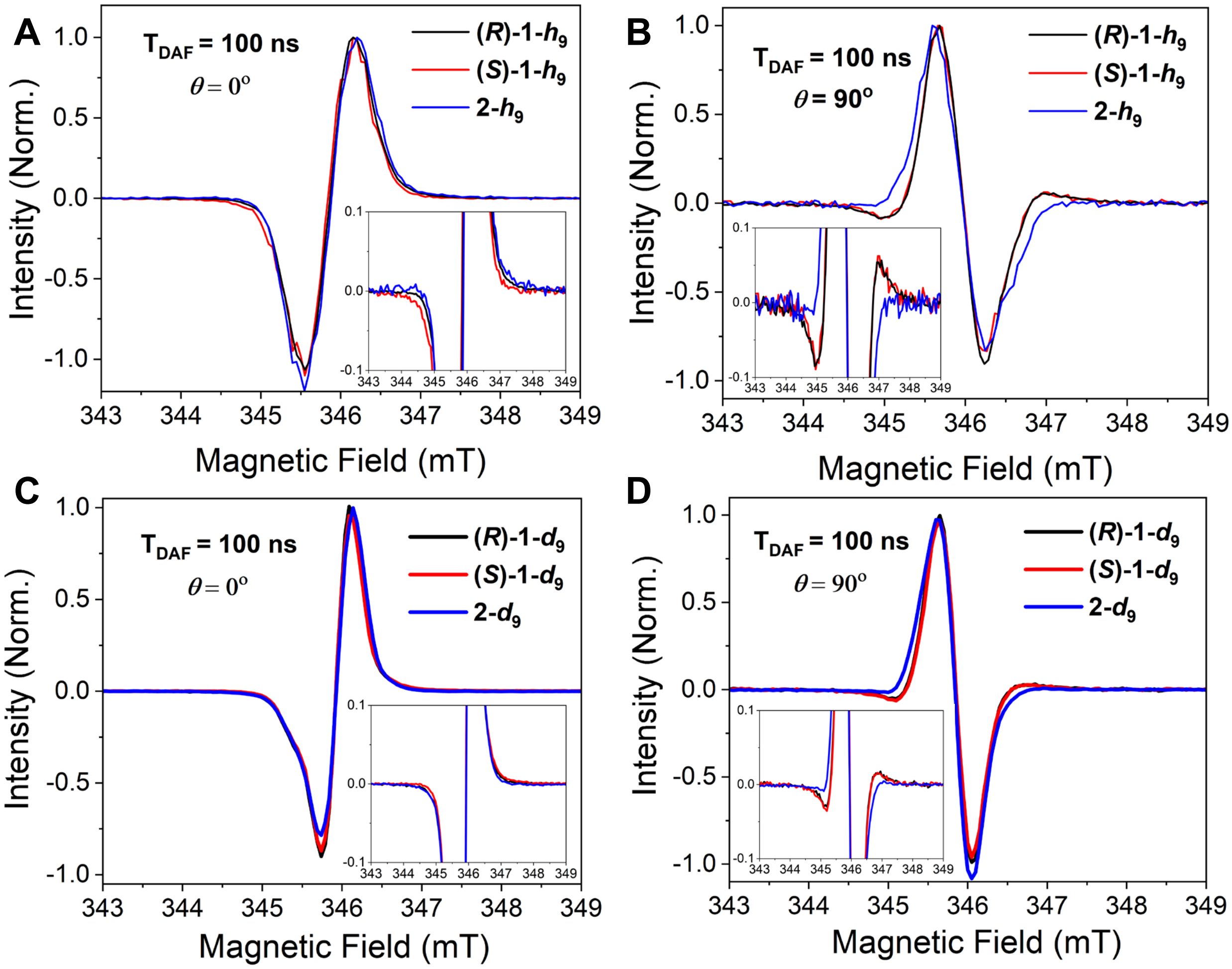

**Fig. 2. TREPR spectra**. TREPR spectra of (*R*)-**1-*h*9**, (*S*)-**1-*h*9**, and **2-*h*9** (**A**, **B**) and (*R*)-**1-*d*9**, (*S*)-**1-*d*9**, and **2-*d*9** (**C**, **D**) oriented in the nematic liquid crystal 5CB at 85 K and 100 ns after a 450 nm, 7 ns laser pulse with the long axis of each molecule (**A**, **C**) 0° and (**B**, **D**) 90° relative to the applied magnetic field direction. Insets: the spectra are shown with their intensities expanded to highlight features characteristic of CISS.

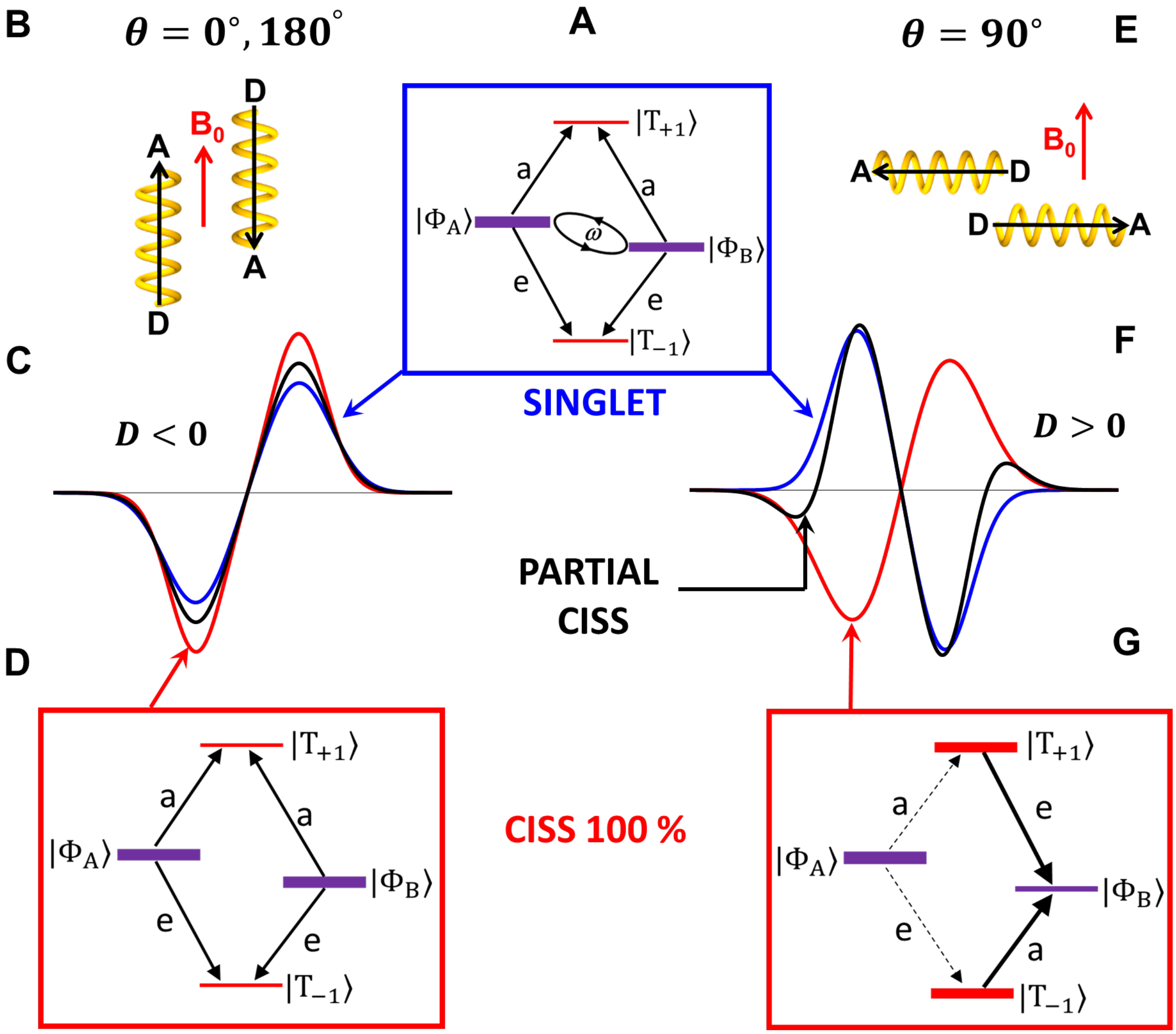

**Fig. 3. The CISS effect on the spin states of SCRPs.** (**A**) SCRP spin states in the absence of CISS and in the presence of a static magnetic field $B_0$ that is much greater than $J$, $D$, and the hyperfine interactions in both radicals for a singlet precursor. The enhanced absorptive (a) and emissive (e) microwave-induced EPR transitions are indicated. (**B**) Schematic of chiral molecules aligned parallel to $B_0$. (**C**) TREPR spectra with $\theta = 0°$ expected for an achiral SCRP (blue trace) and for a chiral SCRP with an initial state having 100% CISS contribution (red trace) or a partial CISS contribution (black trace). (**D**) SCRP spin states for $\theta = 0°$ where the initial state has a 100% CISS contribution. (**E**) Schematic of chiral molecules aligned perpendicular to $B_0$. (**F**) TREPR spectra with $\theta = 90°$ expected for an achiral SCRP (blue trace) and for a chiral SCRP with an initial state having 100% CISS contribution (red trace) or a partial CISS contribution (black trace, rescaled). (**G**) SCRP spin states for $\theta = 90°$ where the initial state has a 100% CISS contribution. The width of the energy levels in (**A**), (**D**), and (**G**) indicates the population of the initial state, while the relative arrow thicknesses in the boxes depict the transition probabilities.

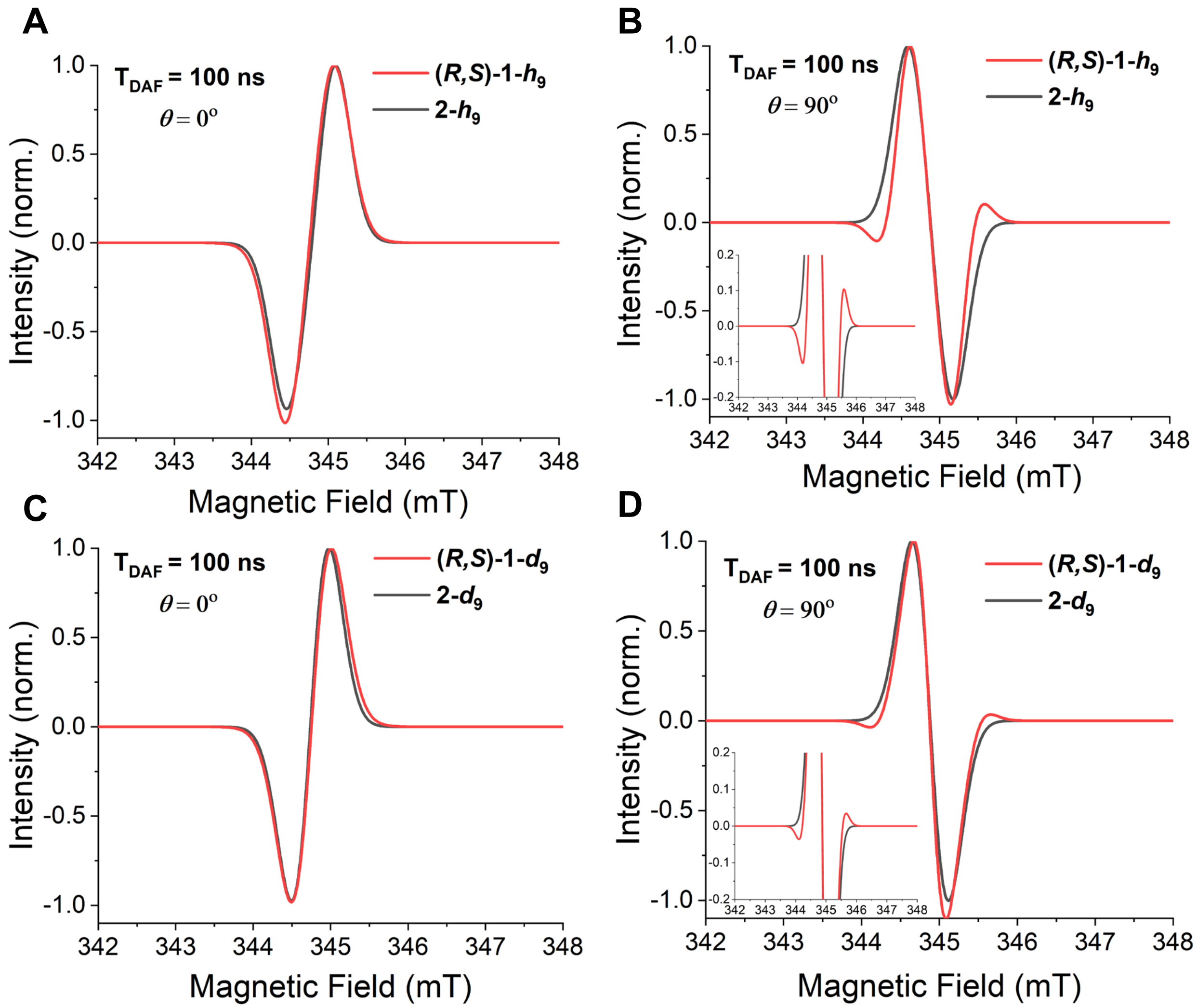

**Fig. 4. Simulations of the TREPR spectra with a minimal model of the SCRP.** The model places one hydrogen nuclear spin-½ on both $PXX^{\bullet+}$ and $NDI^{\bullet-}$ (**A**, **B**) or only on $NDI^{\bullet-}$ (**C**, **D**). The nuclear spins are coupled to each radical with isotropic hyperfine couplings $a_{\mathrm{NDI}} = 6.3$ MHz and $a_{\mathrm{PXX}} = 10$ MHz. Insets: the simulations are shown with their intensities expanded to highlight features characteristic of CISS. The complete list of simulation parameters is given in table S7.

# Supplementary Materials for

## Direct Observation of Chirality-Induced Spin Selectivity in Electron Donor-Acceptor Molecules

Hannah J. Eckvahl,[1†] Nikolai A. Tcyrulnikov,[1†] Alessandro Chiesa,[2†] Jillian M. Bradley,[1] Ryan M. Young,[1] Stefano Carretta,[2*] Matthew D. Krzyaniak,[1*] and Michael R. Wasielewski[1*]

[1]Department of Chemistry, Center for Molecular Quantum Transduction, and Paula M. Trienens Institute for Sustainability and Energy, Northwestern University, Evanston, IL 60208-3113 USA.
[2]Università di Parma, Dipartimento di Scienze Matematiche, Fisiche e Informatiche, Parma, I-43124, Italy.

*Corresponding authors. Email: m-wasielewski@northwestern.edu; mdkrzyanaik@northwestern.edu; stefano.carretta@unipr.it

†These authors contributed equally to this work

**The PDF file includes:**

Materials and Methods
Supplementary Text
Figs. S1 to S12
Tables S1 to S7
References (*50-60*)

## Materials and Methods

Synthesis

The synthetic schemes and structures of the intermediates and final structures are illustrated in Figs. S1-S2. All commercially available materials were used without further purification unless otherwise noted. Characterization of synthesized compounds was performed at Integrated Molecular Structure Education and Research Center (IMSERC) at Northwestern University. Proton and carbon NMR spectra were recorded on a Bruker Avance III 500 MHz system equipped with DCH CryoProbe. Shifts are reported in parts-per-million (ppm) relative to TMS. High Resolution Mass Spectra (HRMS) were obtained with an Agilent LCTOF 6200 series mass spectrometer using electrospray ionization (ESI), APCI, and APPI. CD spectra were obtained with Jasco J-815 CD Instrument Keck Biophysics Facility at Northwestern University.

**NIA-xyl**. In a 250 mL two-neck round bottom flask equipped with a condenser and a dropping funnel, 1,4,5,8-naphthalenetetracarboxylic dianhydride (3.0 g, 11.2 mmol, regenerated prior to use by refluxing overnight in acetic anhydride followed by the removal of the acetic anhydride under reduced pressure) was suspended in 40 mL of dry *N,N*-dimethylformamide. The mixture was heated to reflux under an inert atmosphere and stirred for two hours to ensure the dissolution of the 1,4,5,8-naphthalenetetracarboxylic dianhydride. To this vigorously stirred solution, 2,6-xylidine (677 mg, 5.6 mmol) in 50 ml of dry *N,N*-dimethylformamide was added via a dropping funnel over several hours. The reaction mixture was refluxed overnight under an inert atmosphere, cooled down to room temperature, and the solvent removed under reduced pressure. The crude material was purified by silica gel chromatography (100 % dichloromethane), affording **NIA-xyl** as a brown powder (800 mg, 2.15 mmol, 38 %). $^1$H NMR (500 MHz, $CD_2Cl_2$, δ): 8.76-8.80 (m, 4H), 7.27 (m, 1H), 7.19 (m, 2H), 2.05 (m, 6H). $^{13}$C NMR (126 MHz, $CDCl_3$, δ): 161.43, 160.90, 158.25, 135.00, 134.92, 132.76, 132.47, 132.38, 130.81, 130.60, 128.42, 128.30, 128.26, 127.85,

127.80, 127.04, 126.85, 126.76, 126.14, 122.36, 16.80. ESI-HRMS (*m/z*): calculated for $C_{23}H_9N_5O$ $[M]^+$: 371.0813, found: 371.0803.

**$NH_2$-NDI-xyl**. In a 25 mL round bottom flask equipped with a condenser, **NIA-xyl** (300 mg, 0.81 mmol) was suspended in 10 mL of dry *N,N*-dimethylformamide. The mixture was heated to reflux and stirred for one hour. Hydrazine monohydride (0.5 mL) was added, and the reaction mixture was refluxed for 3 hours. The reaction mixture was cooled to room temperature and the solvent was removed under reduced pressure. The crude material was dissolved in dichloromethane and washed twice with brine. The organic layer was dried over sodium sulfate, filtered, and concentrated under reduced pressure. The crude material was purified by silica gel chromatography (20 % ethyl acetate in dichloromethane), affording **$NH_2$-NDI-xyl** as a yellow solid (270 mg, 0.70 mmol, 87 %). $^1H$ NMR (500 MHz, $CD_2Cl_2$, δ): 8.71-8.76 (m, 4H), 7.26 (m, 1H), 7.18 (d, J = 7.7 Hz, 2H), 5.52 (s, 2H), 2.04 (s, 6H). $^{13}C$ NMR (126 MHz, $CD_2Cl_2$, δ): 161.32, 158.81, 135.00, 132.71, 130.58, 130.34, 128.28, 127.78, 126.64, 126.13, 125.49, 124.83, 16.82. ESI-HRMS (*m/z*): calculated for $C_{22}H_{15}N_3O_4$ $[M+H]^+$: 386.1135, found: 386.1152.

**NMA-NMI-NDI-xyl**. In a 25 mL two-neck round bottom flask equipped with a condenser and a dropping funnel, 5,5′-dimethoxy-1H,1′H,3H,3′H-[6,6′-bibenzo[de]isochromene]-1,1′,3,3′-tetraone(*50*) (327 mg, 0.72 mmol) was suspended in 10 mL of dry *N,N*-dimethylformamide and the mixture was refluxed for one hour under inert atmosphere. To this, **$NH_2$-NDI-xyl** (250 mg, 0.65 mmol) dissolved in 50 mL of dry *N,N*-dimethylformamide was added via a dropping funnel over several hours, and the reaction mixture was further refluxed overnight under an inert atmosphere. The reaction mixture was cooled to room temperature and the solvent was removed under reduced pressure. Due to purification difficulties and the tendency of the product to undergo hydrolysis of the anhydride moiety, the crude material was passed through a silica gel pad (10 % ethyl acetate in dichloromethane) and used without further purification right away. A small fraction of the crude was purified using preparatory TLC for characterization. $^{1}$H NMR (500 MHz, $CD_2Cl_2$, δ): 8.84 (m, 4H), 8.55 (s, 1H), 8.47-8.51 (m, 2H), 8.43 (d, J = 6.8 Hz, 1H), 7.51-7.57 (m, 2H), 7.28 (m, 1H), 7.21 (d, J = 7.6 Hz, 2H), 3.87 (s, 3H), 3.84 (s, 3H), 2.08 (s, 6H). $^{13}$C NMR (126 MHz, $CD_2Cl_2$, δ): 162.00, 160.53, 160.44, 159.92, 159.86, 156.06, 155.87, 133.44, 132.61, 132.48, 132.22, 132.08, 132.05, 131.46, 131.43, 131.32, 130.47, 129.11, 128.58, 128.29, 128.00, 127.81, 127.56, 127.48, 126.41, 126.10, 125.93, 125.46, 124.25, 123.76, 122.09, 120.58, 119.01, 118.72, 117.95, 29.68, 17.59. APCI-HRMS (*m/z*): calculated for $C_{48}H_{27}N_3O_{11}$ $[M]^+$: 821.1651, found: 821.1627.

**Br-Ph-(NMI)$_2$-NDI-xyl**. In a 50 mL round bottom flask equipped with a condenser, **NMA-NMI-NDI-xyl** (210 mg, crude), 4-bromoaniline (132 mg, 0.77 mmol), and zinc (II) acetate dihydrate (15 mg) were dissolved in 20 mL of dry *N,N*-dimethylformamide and the mixture was refluxed overnight under an inert atmosphere. The reaction mixture was cooled to room temperature and the solvent was removed under reduced pressure. The crude material was purified by silica gel chromatography (gradient from 100 % dichloromethane to 5 % ethyl acetate in dichloromethane), affording **Br-Ph-(NMI)$_2$-NDI-xyl** as a brown solid (40 mg, 0.30 mmol). $^{1}$H NMR (500 MHz, $CD_2Cl_2$, δ): 8.80-8.90 (m, 4H), 8.56 (s, 1H), 8.50 (m, 2H), 8.43 (dd, $J_1$ = 7.0 Hz, $J_2$ = 1.1 Hz, 1H), 7.66 (d, J = 8.6 Hz, 2H), 7.49-7.58 (m, 3H), 7.45 (dd, $J_1$ = 8.6 Hz, $J_2$ = 1.1 Hz, 1H), 7.29 (m, 1H), 7.18-7.23 (m, 4H), 3.86 (m, 6H), 2.08 (s, 6H). $^{13}$C NMR (126 MHz, $CD_2Cl_2$, δ): 163.20, 163.01, 161.25, 159.99, 159.96, 159.18, 159.11, 155.13, 155.07, 135.02, 135.01, 134.05, 132.66, 131.99, 131.72, 131.64, 131.57, 131.33, 131.30, 130.72, 130.70, 130.49, 129.87, 129.70, 128.69, 128.36, 127.83, 127.19, 127.13, 127.03, 126.75, 126.70, 125.64, 125.44, 123.90, 123.80, 123.48, 123.34, 122.74, 122.08, 121.75, 121.24, 117.19, 116.37, 28.93, 16.85. ESI-HRMS (*m/z*): calculated for $C_{54}H_{31}BrN_4O_{10}Na$ $[M+Na]^+$: 999.1109, found: 999.1125.

**(*R*,*S*)-1-*d*$_9$**. In a 25 mL round bottom two-neck flask equipped with a condenser and a septum were combined **Br-Ph-(NMI)$_2$-NDI-xyl** (20 mg, 0.02 mmol), PXX-*d*$_9$-Bpin (Fig. S2) (17 mg, 0.04 mmol), [1,1′-bis(diphenylphosphino)ferrocene]dichloropalladium(II) (3 mg, 0.004 mmol), and potassium carbonate (20 mg). The flask was evacuated and filled with nitrogen three times. To this mixture, 10 mL of a degassed mixture of toluene, ethanol, and water (2/1/1 vol. ratio) was added through the septum. The reaction mixture was heated at 110 $^0$C overnight under an inert atmosphere. The reaction mixture was cooled to room temperature, dissolved in dichloromethane, and washed twice with brine. The organic layer was dried over sodium sulfate, filtered, and concentrated under reduced pressure. The crude material was purified by silica gel chromatography (gradient from 100 % dichloromethane to 5 % ethyl acetate in dichloromethane), affording **(*R*,*S*)-1-*d*$_9$** as a yellow solid (17 mg, 0.014 mmol, 72 %). $^1$H NMR (500 MHz, $CD_2Cl_2$, δ): 8.81-8.88 (m, 4H), 8.58 (s, 1H), 8.55 (s, 1H), 8.51 (dd, $J_1$ = 6.5 Hz, $J_2$ = 1.5 Hz, 1H), 8.47 (dd, $J_1$ = 7.0 Hz, $J_2$ = 1.0 Hz, 1H), 7.62 (d, J = 8.4 Hz, 2H), 7.52-7.60 (m, 3H), 7.47 (dd, $J_1$ = 8.5 Hz, $J_2$ = 1.0 Hz, 1H), 7.41 (d, J = 8.2 Hz, 2H), 7.29 (m, 1H), 7.21 (d, J = 7.7 Hz, 2H), 3.89 (s, 3H), 3.87 (s, 3H), 2.09 (s, 6H). $^{13}$C NMR (126 MHz, $CD_2Cl_2$, δ): 163.48, 163.29, 161.26, 160.02, 159.98, 159.19, 159.12, 155.16, 155.09, 151.74, 151.60, 143.78, 143.36, 139.24, 135.03, 134.10, 132.66, 132.03, 131.65, 131.34, 131.31, 130.72, 130.47, 130.38, 129.72, 129.41, 128.65, 128.35, 128.25, 127.83, 127.20, 127.14, 127.03, 126.76, 125.65, 125.55, 124.03, 123.77, 123.50, 123.40, 122.73, 122.30, 121.25, 121.09, 120.70, 117.22, 116.36, 110.96, 110.56, 28.93, 16.86. APPI-HRMS (*m/z*): calculated for $C_{74}H_{31}D_9N_4O_{12}$ $[M]^+$: 1185.3208, found: 1185.3225.

**(*R*,*S*)-1-*h*$_9$**. Using a procedure described for the synthesis of (*R,S*)-**1-*d9***, (*R,S*)-**1-*h9*** was obtained from 20 mg (0.02 mmol) of **D** and 18 mg of PXX-Bpin(*31*) as a yellow solid (16 mg, 66 %). $^1$H NMR (500 MHz, $CD_2Cl_2$, δ): 8.81-8.88 (m, 4H), 8.58 (s, 1H), 8.55 (s, 1H), 8.51 (dd, $J_1$ = 6.6 Hz, $J_2$ = 1.4 Hz, 1H), 8.47 (dd, $J_1$ = 7.0 Hz, $J_2$ = 0.7 Hz, 1H), 7.63 (d, J = 8.2 Hz, 2H), 7.50-

7.59 (m, 4H), 7.46 (dd, $J_1$ = 8.2 Hz, $J_2$ = 0.8 Hz, 1H), 7.41 (d, J = 8.2 Hz, 2H), 7.26-7.33 (m, 2H), 7.21 (d, J = 7.6 Hz, 2H), 7.16 (d, J = 7.8 Hz, 1H), 7.05-7.10 (m, 2H), 6.92 (dd, $J_1$ = 9.1 Hz, $J_2$ = 3.7 Hz, 2H), 6.74 (d, J = 7.8 Hz, 1H), 6.62 (dd, $J_1$ = 5.6 Hz , $J_2$ = 1.8 Hz, 1H), 3.89 (s, 3H), 3.87 (s, 3H), 2.09 (s, 6H). $^{13}$C NMR (126 MHz, $CD_2Cl_2$, δ): 163.47, 163.28, 161.26, 160.02, 159.99, 159.19, 159.13, 155.22, 155.15, 151.83, 151.68, 143.87, 143.46, 139.29, 135.05, 135.03, 134.15, 132.70, 132.08, 131.70, 131.62, 131.51, 131.33, 131.30, 130.71, 130.37, 129.71, 129.42, 128.65, 128.50, 128.36, 128.27, 127.83, 127.58, 127.20, 127.13, 127.06, 126.81, 126.80, 126.74, 126.48, 125.84, 125.69, 125.62, 124.38, 124.09, 123.84, 123.55, 123.45, 122.78, 122.37, 121.31, 121.12, 120.75, 119.47, 117.28, 116.81, 116.55, 116.43, 111.00, 110.61, 108.07, 107.94, 28.93, 16.84. ESI-HRMS (*m/z*): calculated for $C_{74}H_{41}N_4O_{12}$ $[M+H]^+$: 1177.2715, found: 1177.2707.

**PXX-*d*$_9$-Br.** Under a nitrogen atmosphere, **PXX-*d*$_{10}$** (*42*) (1 eq, 0.34 mmol, 100 mg) and *N*-bromosuccinimide (1 eq, 0.34 mmol, 60.9 mg) were suspended in dry dimethylformamide (6.5 mL). The mixture was sparged with nitrogen for 15 minutes, then stirred at room temperature for 16 hr. The mixture was poured into water (60 mL) and vacuum filtered. The yellow solid was washed three times with water (20 mL), then dried in a vacuum oven overnight to yield **PXX-*d9*-Br** as a dull yellow solid (104 mg). The product was used without further purification.

**PXX-*d*$_9$-Bpin.** Under a nitrogen atmosphere, **PXX-*d*$_9$-Br** (1 eq, 0.28 mmol 103 mg), bis(pinacolato)diboron (3 eq, 0.83 mmol, 212 mg), potassium acetate (6 eq, 1.67 mmol, 164 mg), and [1,1′-bis(diphenylphosphino)ferrocene]dichloropalladium(II) (0.05 eq, 0.014 mmol, 10.2 mg) were dissolved in dioxanes (15 mL, stored in an inert atmosphere over molecular sieves). The mixture was sparged with nitrogen for 15 minutes, then stirred at reflux 12 hr. After cooling to room temperature, the solvent was removed under reduced pressure. The product was purified with flash chromatography (50/50 hexanes/dichloromethane mobile phase), yielding **PXX-*d*$_9$-Bpin** as a fluorescent yellow solid (50 mg, 43%). $^1$H NMR (500 MHz, $CDCl_3$, δ): 1.37. $^{13}$C NMR (126

MHz, $CDCl_3$, δ): 155.43, 152.63, 144.22, 143.75, 135.20, 131.25, 121.47, 121.35, 111.91, 111.28, 83.51, 24.98. ESI-HRMS (*m/z*): calculated for $C_{26}H_{12}D_9BO_4$ $[M]^+$: 417.2098, found: 417.2120.

**PXX-*d*9-xy-Br.** Under a nitrogen atmosphere, **PXX-*d*9-Bpin** (1 eq, 0.083 mmol 34.2 mg), 2,5-dibromo-*p*-xylene (20 eq, 1.64 mmol, 433 mg), potassium carbonate (4 eq, 0.32 mmol, 45.3 mg), and tetrakis(triphenylphosphine)palladium(0) (0.05 eq, 0.004 mmol, 4.8 mg) were dissolved in dioxanes (10 mL, distilled, then stored in an inert atmosphere over molecular sieves) and water (1 mL). The mixture was sparged with nitrogen for 15 minutes, then stirred at reflux 12 hr. After cooling to room temperature, the solvent was removed under reduced pressure. The product was purified with flash chromatography (90/10 hexanes/dichloromethane mobile phase), yielding **PXX-*d*9-xy-Br** as a fluorescent yellow solid (31 mg, 80%). $^1H$ NMR (500 MHz, $CDCl_3$, δ): 7.47 (s, 1H), 7.05 (s, 1H), 2.38 (s, 3H), 2.01 (s, 3H). $^{13}C$ NMR (126 MHz, $CDCl_3$, δ): 152.58, 152.10, 144.44, 144.18, 138.27, 136.23, 135.15, 133.55, 132.52, 131.38, 131.18, 129.48, 123.83, 121.56, 121.52, 111.76, 111.39, 22.31, 19.14. ESI-HRMS (*m/z*): calculated for $C_{28}H_8D_9BrO_2$ $[M]^+$: 473.0977, found: 473.0990.

**PXX-*d*9-xy-Bpin.** Under a nitrogen atmosphere, **PXX-*d*9-xy-Br** (1 eq, 0.046 mmol 21.5 mg), bis(pinacolato)diboron (2 eq, 0.092 mmol, 23.4 mg), potassium acetate (4 eq, 0.18 mmol, 18 mg), and [1,1′-bis(diphenylphosphino)ferrocene]dichloropalladium(II) (0.05 eq, 0.0023 mmol, 1.7 mg) were dissolved in dioxanes (5 mL, distilled, then stored in an inert atmosphere over molecular sieves). The mixture was sparged with nitrogen for 15 minutes, then stirred at reflux 12 hr. After cooling to room temperature, the solvent was removed under reduced pressure. The product was purified with flash chromatography (70/30 hexanes/dichloromethane mobile phase), yielding **PXX-*d*9-xy-Bpin** as a fluorescent yellow solid (18 mg, 75%). $^1H$ NMR (500 MHz, $CDCl_3$, δ): 7.69 (s, 1H), 7.00 (s, 1H), 2.52 (s, 3H), 2.03 (s, 3H), 1.36 (s, 12H). $^{13}C$ NMR (126 MHz, $CDCl_3$, δ): 152.64, 151.90, 144.37, 144.21, 142.28, 141.58, 137.55, 132.94, 132.60, 131.78, 131.17, 129.48,

121.60, 121.46, 111.63, 111.50, 88.26, 83.47, 29.71, 24.93, 24.92, 21.66, 19.11. ESI-HRMS (*m/z*): calculated for $C_{34}H_{20}D_9BO_4$ $[M]^+$: 521.2724, found: 521.2755.

**Br-NMI-NDI-dtbp. NIA-dtbp** and **Br-NMI-NH₂** were synthesized as previously reported.(*38, 51*) Under a nitrogen atmosphere, Br-NMI-$NH_2$ (1 eq, 0.25 mmol, 100 mg) and NIA-dtbp (1.5 eq, 0.38 mmol, 173 mg) were dissolved in ethanol (5 mL, 100%). The mixture was sparged with nitrogen for 15 minutes, then stirred at reflux 12 h (overnight). The precipitate was isolated by vacuum filtration while the mixture was still hot, yielding **Br-NMI-NDI-dtbp** as a white powder (142 mg, 67 %). $^1$H NMR (500 MHz, $CDCl_3$, δ): 8.89-8.84 (m, 4H), 8.37 (dd, $J_1$ = 7.3 Hz, $J_2$ = 1.2 Hz, 1H), 8.66 (dd, $J_1$ = 8.6 Hz, $J_2$ = 1.2 Hz, 1H), 8.49 (d, J = 7.8 Hz, 1H), 8.11 (d, J = 7.9 Hz, 1H), 7.91 (dd, $J_1$ = 8.6 Hz, $J_2$ = 7.2 Hz, 1H), 7.60 (d, J = 8.7 Hz, 1H), 7.48 (dd, $J_1$ = 8.6 Hz, $J_2$ = 2.3 Hz, 1H), 7.28-7.21 (m, 2H), 7.00 (dd, $J_1$ = 7.3 Hz, $J_2$ = 2.2 Hz, 1H), 2.21-2.26 (m, 6H), 1.33-1.30 (m, 9H), 1.27 (s, 9H). $^{13}$C NMR (126 MHz, $CDCl_3$, δ): 163.84, 163.24, 163.21, 162.42, 150.39, 143.68, 143.66, 135.16, 134.84, 134.79, 134.27, 133.81, 132.59, 131.99, 131.97, 131.75, 131.54, 131.29, 131.23, 130.97, 130.90, 130.88, 129.52, 129.03, 129.01, 128.27, 127.49, 127.46, 127.43, 127.32, 127.29, 127.16, 126.93, 126.73, 123.12, 122.23, 35.57, 34.28, 31.74, 31.19, 17.40. APPI-HRMS (*m/z*): calculated for $C_{48}H_{38}BrN_3O_6Na$ $[M+Na]^+$: 854.1842, found: 854.1827.

**2-*d*₉.** Under a nitrogen atmosphere, **PXX-*d*₉-xy-Bpin** (1.22 eq, 0.029 mmol 15 mg), **Br-NMI-NDI-dtbp** (1 eq, 0.024 mmol, 19.7 mg), potassium carbonate (2.5 eq, 0.059 mmol, 8.2 mg), and [1,1′-bis(diphenylphosphino)ferrocene]dichloropalladium(II) (0.1 eq, 0.0023 mmol, 1.7 mg) were dissolved in dioxanes (5 mL, distilled, then stored in an inert atmosphere over molecular sieves) and water (0.5 mL, distilled). The mixture was sparged with nitrogen for 15 minutes, then stirred at reflux 12 h (overnight). After cooling to room temperature, the solvent was removed under reduced pressure. The product was purified with flash chromatography (dichloromethane eluent), yielding **2-*d9*** as a yellow solid (14 mg, 42%). $^1$H NMR (500 MHz, $CDCl_3$, δ): 8.90-8.81 (m, 4H), 8.78-8.69 (m, 2H), 8.08 (t, J = 8.0 Hz, 1H), 7.80 (t, J = 7.8 Hz, 1H), 7.62 (t, J = 8.4 Hz, 1H), 7.60 (d, J = 8.7 Hz, 1H), 7.48 (dd, $J_1$ = 8.6 Hz, $J_2$ = 2.3 Hz, 1H), 7.31 (d, J = 3.0 Hz, 1H), 7.24-7.19 (m, 2H), 7.02 (dd, $J_1$ = 13.6 Hz, $J_2$ = 2.2 Hz, 1H), 2.27-2.18 (m, 6H), 2.13 (s, 3H), 2.07 (s, 3H), 1.34-1.31 (m, 9H), 1.28 (s, 9H). $^{13}$C NMR (126 MHz, $CDCl_3$, δ): 163.92, 163.86, 163.73, 162.39, 152.60, 152.14, 150.40, 147.28, 147.27, 147.23, 147.21, 144.48, 144.25, 144.23, 143.75, 143.73, 139.53, 139.49, 137.54, 137.48, 135.58, 135.24, 135.22, 134.73, 134.67, 134.45, 134.44, 134.40, 134.39, 134.20, 133.64, 133.62, 133.58, 133.57, 133.25, 133.18, 132.24, 132.08, 132.06, 132.01, 131.96, 131.71, 131.69, 131.61, 131.59, 131.50, 131.48, 131.35, 131.34, 131.32, 131.22, 130.93, 130.87, 130.82, 129.67, 129.59, 129.03, 129.00, 128.97, 128.12, 128.07, 127.57, 127.54, 127.50, 127.45, 127.35, 127.31, 127.17, 127.10, 127.07, 127.05, 126.70, 123.07, 123.04, 121.96, 121.94, 121.91, 121.63, 121.59, 111.90, 111.83, 111.44, 111.40, 35.60, 34.32, 34.30, 31.77, 31.23, 31.21, 29.71, 22.70, 19.69, 19.66, 19.48, 19.47, 17.54, 17.48, 17.47, 17.45, 14.20, 14.12. APPI-HRMS (*m/z*): calculated for $C_{76}H_{46}D_9N_3O_8Na$ [M+Na]$^+$: 1169.4452, found: 1169.4423.

**2-*h*$_9$.** Under a nitrogen atmosphere, **PXX-xy-Bpin** (1.5 eq, 0. 222 mmol, 112.6 mg), **Br-NMI-NDI-dtbp** (1 eq, 0.150 mmol, 124.8 mg), and potassium carbonate (2 eq, 0.300 mmol, 41.5 mg) were dissolved in tetrahydrofuran (15 mL) and water (1.5 mL). The mixture was sparged with nitrogen for 15 minutes. The flask was opened briefly to add tetrakis(triphenylphosphine) palladium(0) (0.05 eq, 0.0075 mmol, 8.7 mg), then sparged with nitrogen for another 15 minutes. The mixture was then heated to reflux and stirred 12 h (overnight). After cooling to room temperature, the solvent was removed under reduced pressure. The product was purified with flash chromatography (using 3% ethyl acetate in dichloromethane as the mobile phase), yielding **2-*h*$_9$** as a yellow-green solid (100 mg, 59 %). $^1$H NMR (500 MHz, $CDCl_3$, δ): 8.90-8.82 (m, 4H), 8.76 (dt,$J_1$ = 7.4 Hz, $J_2$ = 3.0 Hz, 1H), 8.72 (t,J = 7.7 Hz, 1H), 8.07 (t, J = 7.9 Hz, 1H), 7.82-7.74 (m,

2H), 7.60 (d, J = 8.7 Hz, 1H), 7.48 (dd,$J_1$ = 8.7 Hz, $J_2$ = 2.3 Hz, 1H), 7.35 (dd,$J_1$ = 9.1 Hz, $J_2$ = 1.9 Hz, 1H), 7.26-7.19 (m, 3H), 7.14-6.96 (m, 6H), 6.93 (dd, $J_1$ = 16.6 Hz, $J_2$ = 9.2 Hz, 1H), 6.78 (dd, $J_1$ = 7.8 Hz, $J_2$ = 2.4 Hz, 1H), 6.67 (m, 1H), 2.27-2.20 (m, 6H), 2.13 (s, 3H), 2.07 (s, 3H), 1.34-1.32 (m, 9H), 1.28 (s, 9H). $^{13}$C NMR (126 MHz, $CDCl_3$, δ): 163.93, 163.92, 163.86, 163.74, 163.73, 162.40, 152.65, 152.64, 152.18, 150.40, 147.28, 147.24, 144.54, 144.54, 144.31, 144.29, 143.75, 143.74, 139.55, 139.51, 137.54, 137.49, 135.58, 135.58, 135.26, 135.23, 134.74, 134.69, 134.47, 134.45, 134.41, 134.40, 134.20, 133.64, 133.64, 133.59, 133.58, 133.28, 133.27, 133.19, 133.18, 132.27, 132.25, 132.15, 132.10, 132.06, 131.73, 131.72, 131.61, 131.58, 131.57, 131.51, 131.49, 131.46, 131.38, 131.36, 131.35, 131.33, 130.94, 130.88, 130.82, 129.75, 129.68, 129.04, 129.01, 128.98, 128.12, 128.08, 127.84, 127.72, 127.57, 127.54, 127.50, 127.45, 127.35, 127.31, 127.24, 127.18, 127.10, 127.06, 126.70, 126.61, 126.58, 125.31, 125.21, 123.07, 123.03, 121.96, 121.93, 121.91, 121.63, 121.59, 120.26, 120.24, 117.52, 117.39, 117.38, 111.92, 111.84, 111.45, 111.42, 108.78, 108.64, 108.56, 35.60, 34.32, 34.30, 31.77, 31.23, 31.22, 29.71, 19.70, 19.66, 19.48, 19.47, 17.54, 17.48, 17.47, 17.45. APPI-HRMS (*m/z*): calculated for $C_{76}H_{55}N_3O_8$ ESI-HRMS (*m/z*): calculated for $C_{76}H_{55}N_3O_8$ $[M]^+$: 1137.3984, found: 1137.3998.

<u>Optical Spectroscopic Characterization</u>

The two enantiomers of (*R,S*)-**1-*h*9** and (*R,S*)-**1-*d*9** were separated using a chiral HPLC column (Fig. S3). Their steady-state UV-Vis spectra were obtained with a Shimadzu-1800 spectrophotometer in toluene. The spectrum of racemic (*R,S*)-**1-*h*9** along with the spectra of PXX-*h*9, racemic (*R,S*)-$NMI_2$, and NDI are shown in Fig. S4a. The spectrum of (*R,S*)-**1-*h*9** is approximately an additive combination of those of PXX, (*R,S*)-$NMI_2$, and NDI, thus indicating that the electronic coupling between these redox components is relatively weak. The corresponding spectra of (*R,S*)-**1-*d*9** are nearly identical (Fig. S4c). Circular dichroism (CD) spectra of each enantiomer show the expected mirror image symmetry over the absorption range of the chiral $NMI_2$ bridge molecule (Figs. S4b and S4d), which is in full agreement with the CD spectra reported for $NMI_2$ previously.(*32*)

Details of the transient absorption instrumentation have been described previously.(*52*) Briefly, ~50% of the output of a 1 kHz amplified Ti:sapphire system at 827 nm (1 W, 100 fs, Spitfire, Spectra Physics) is used to pump a non-collinear optical parametric amplifier (TOPAS-Prime, Light-Conversion, LLC.) tuned to generate the ~120 fs, 520 nm pump pulses. The pump is depolarized to minimize polarization-specific dynamics. In the nsTA experiment, the probe is generated in a separately delayed broadband laser system (EOS, Ultrafast Systems, LLC). The transmitted probe is detected on a commercial spectrometer (customized Helios-EOS, Ultrafast Systems, LLC).

Transient absorption (TA) measurements were used to obtain the photoinduced electron transfer dynamics for r=the racemic mixtures (*R* and *S*)-**1-*h*₉**, (*R* and *S* )-**1-*d*₉** as well as the achiral reference molecules **2-*h*₉** and **2-*d*₉** in a butyronitrile (PrCN) glass at 105 K using a 120-fs, 450 nm pump pulse and monitoring the transient spectra at 100 fs-250 μs (Fig. S5). These data were globally fit over the entire time range to yield evolution-associated spectra (Fig. S6). Photoexcitation of (*R* and *S*)-**1-*h*₉** results in the ground state bleach (GSB) of its 450 nm absorption accompanied by the appearance of the excited state absorption (ESA) bands of the lowest excited singlet state of PXX-$h_9$ ($^{1*}$PXX) at 630, 779, and 1177 nm, as well as stimulated emission features at 483 and 523 nm (Figs. 5a and S6a). These transient features decay in $\tau_{CS1}$ = 12.4 ± 0.4 ps accompanied by the rise of a complex transient signal showing a prominent 979 nm band characteristic of PXX-$h_9^{\bullet+}$.(*31*) The 410 nm absorption characteristic of $NMI^{\bullet-}$ is largely cancelled out by the PXX-$h_9$ GSB.(*53*) A second electron transfer producing $NDI^{\bullet-}$occurs in $\tau_{CS2}$ = 203 ± 8 ps as indicated by the formation of the sharp features at 474 and 601 nm.(*53*) Further low amplitude absorption changes of PXX-$h_9^{\bullet+}$ and $NDI^{\bullet-}$ occur in $\tau_{rlx}$ = 32 ± 5 ns, which are attributed to a relaxation process in the low temperature glassy solvent followed by decay of the PXX-$h_9^{\bullet+}$-NDI$^{\bullet-}$ SCRP to its ground state in $\tau_{CR}$ = 65.9 ± 0.7 μs. The corresponding TA data for (*R* and *S*)-**1-*d*₉** (Figs. S5c and S6c) exhibit spectra and kinetics very similar to those of (*R* and *S*)-**1-*h*₉**, where the charge recombination time of the PXX-$d_9^{\bullet+}$-NDI$^{\bullet-}$ SCRP to its ground state $\tau_{CR}$ = 51.1 ± 0.3 μs.

TA measurements under the same conditions were conducted on the achiral reference molecule **2-*h*₉**. The transient spectra at selected times are shown in Fig. S5b. The spectral features are very similar to those of (*R* and *S*)-**1-*h*₉**, and the time constants for charge separation and recombination obtained from the global fits to the data are $\tau_{CS1}$ = 4.8 ± 0.3 ps, $\tau_{CS2}$ = 213 ± 6 ps, $\tau_{rlx}$ = 4.8 ± 0.5 ns, and $\tau_{CR}$ = 46.4 ± 0.6 μs. The TA data for achiral **2-*d*₉** (Figs. S5d and S6d) exhibit spectra and kinetics very similar to those of **1-*h*₉**, where the charge recombination time of the PXX-$d_9^{\bullet+}$-NDI$^{\bullet-}$ SCRP to its ground state in $\tau_{CR}$ = 57.0 ± 0.1 μs. The TA data show that (*R* and *S*)-**1** and **2** have ultrafast charge separation rates, which ensures an initial singlet state precursor, as well as long SCRP recombination lifetimes, making them excellent candidates for studying the CISS effect on their spin dynamics.

Transient Electron Paramagnetic Resonance (TREPR) Spectroscopy

All EPR measurements were performed at X-band (~9.6 GHz) on a Bruker Elexsys E680 X/W EPR spectrometer equipped with a split ring resonator (ER4118X-MS3). The temperature was maintained at 85 K using an Oxford Instruments CF935 continuous-flow cryostat using liquid nitrogen. The sample was photoexcited at 450 nm with 7 ns, 1 mJ laser pulses generated via an optical parametric oscillator (Spectra-Physics BasiScan) pumped with the 355 nm output of a frequency-tripled Nd:YAG laser (Spectra-Physics Quanta-Ray Lab-170+-10H) operating at a repetition rate of 10 Hz. The laser light was coupled into the resonator via a fiber optic (Thorlabs FT1000UMT) and collimator placed outside the cryostat window.

All EPR measurements were performed in the liquid crystal, 5CB (Sigma-Aldrich). Solutions with an O.D. of 0.5 over a 2 mm pathlength were loaded into quartz tubes (2.40 mm o.d., 2.00 mm i.d.), subjected to three freeze-pump-thaw cycles on a vacuum line ($10^{-3}$ Torr), and sealed with a hydrogen torch. The samples slightly warmed, then aligned in a 1 T magnetic field for 5 minutes and quickly loaded into the already cold resonator at 85 K. Two sample orientations were collected, one, as loaded, with the 5CB director parallel to the magnetic field, and other by rotating the sample 90 degrees to generate a liquid crystal director perpendicular to the magnetic field.

Transient continuous-wave EPR (TREPR) spectra were collected following photoexcitation, the kinetic traces of the transient magnetization at a fixed magnetic field were acquired in quadrature under CW irradiation (0.62 mW). Sweeping the magnetic field gave 2D spectra of magnetic field versus time. For each kinetic trace, the average of the signal acquired prior to the laser pulse was subtracted from the data. The kinetic traces recorded at magnetic field values off-resonance were considered background signals, whose average was also subtracted from all kinetic traces. The spectra were then smoothed in time domain using a Savitzky-Golay filter and appropriately phased into absorption/emission and dispersion. All processing and fitting of the spectra were performed in Matlab using home written scripts and the simulation package EasySpin v6.0-dev.(*54, 55*)

Out-of-phase electron spin echo envelope modulation (OOP-ESEEM) spectroscopy was collected 50 ns following photoexcitation using a two-pulse, $\pi/2 - \tau - \pi$, pulse sequence. Microwave pulses were generated using a 1 kW TWT amplifier (Applied Systems Engineering 117X) in a partially over-coupled resonator to minimize ringing. Using pulse lengths of $\pi/2$ = 16 ns and $\pi$ = 32 ns, the pulse sequence was started at $\tau$ = 100 ns and incremented $\Delta\tau = 16$ ns, a standard 4-step phase cycle was utilized and the integrated spin echo intensity was recorded.

The time domain OOP-ESEEM traces were fit with the equation:(*38*)

$$S(\tau) = e^{-\tau/T_d} \int_0^{\pi} \sin\big(J + D(\theta)\big)\, \tau\, W(\theta) \sin\theta \, d\theta \qquad \text{(S1)}$$

where $T_d$ is a relaxation time, $J$ is the exchange coupling and $D(\theta)$ is the dipolar coupling and $W(\theta)$ is the weighting function ordering imposed by the liquid crystal. The dipolar coupling takes the form:

$$D(\theta) = \frac{\mu_0}{4\pi h} \frac{g_D g_A \beta_e^2}{r_{DA}^3} (1 - 3cos^2\theta) \qquad \text{(S2)}$$

When calculating the distance, $g_D$ and $g_A$ were both taken as the free electron g-value, $g_e = 2.0023$. In the main text we have also introduced the constant $d \equiv \frac{\mu_0}{4\pi h} \frac{g_D g_A \beta_e^2}{r_{DA}^3} = 52.04\ \text{MHz nm}^3/r_{DA}^3$. The parallel and perpendicular spectra were fit simultaneously with weighting function $W(\theta) = e^{\lambda(3\cos^2\theta - 1)/2}$, where $\lambda = 5$ for the parallel spectra and $\lambda = -5$ for the perpendicular spectra, resulting in an orientation distribution with a FWHM of ~20°. We have checked that this orientation distribution does not produce a visible effect on the resulting TREPR spectra.

The OOP-ESEEM data and fits are shown in Figs. S7 and S8 and the fitting parameters are summarized in Table S1.

<u>Density Functional Theory Calculations</u>

Geometry optimizations of (*R*)-**1-*h*₉**, (*S*)-**1-*h*₉**, and **2-*h*₉** were performed using density functional theory (DFT) at B3LYP/6-31G* level of theory in QChem (version 5.1).(*56*) The optimized structures are shown in Fig. S9 and the optimized cartesian coordinates are given in Tables S2-S4.

**Supplementary Text**

<u>Spin-Correlated Radical Pair (SCRP) Theory</u>

*SCRP with no CISS Effect*

The radical pair spin Hamiltonian is:

$$\hat{\mathcal{H}} = \beta_e \boldsymbol{B}_0 \boldsymbol{g}_D \hat{\boldsymbol{S}}_D + \beta_e \boldsymbol{B}_0 \boldsymbol{g}_A \hat{\boldsymbol{S}}_A + J \hat{\boldsymbol{S}}_D \hat{\boldsymbol{S}}_A + \hat{\boldsymbol{S}}_D \boldsymbol{D}_{DA} \hat{\boldsymbol{S}}_A + \hat{\mathcal{H}}_{HF} \qquad \text{(S3)}$$

The first two terms are the electron Zeeman interaction with the static magnetic field, $\boldsymbol{B}_0$; $J$ is the exchange coupling; $\boldsymbol{D}_{DA}$ is the electron-electron dipole interaction matrix with principal elements of $[D \quad D \quad -2D]$ and $D$ follows the definition given in eqn S2; and within the final term, $\widehat{\mathcal{H}}_{HF}$, are the electron-nuclear hyperfine interactions.

In the coupled spin basis, $|T_+\rangle, |T_0\rangle, |T_-\rangle, |S\rangle$, this leads to the Hamiltonian matrix:(*39, 40, 57*)

$$\widehat{\mathcal{H}} = \begin{bmatrix} \upsilon_+ + \frac{J}{4} + \frac{D}{4} & 0 & 0 & 0 \\ 0 & \frac{J}{4} - \frac{D}{2} & 0 & \upsilon_- \\ 0 & 0 & -\upsilon_+ + \frac{J}{4} + \frac{D}{4} & 0 \\ 0 & \upsilon_- & 0 & -\frac{3J}{4} \end{bmatrix} \tag{S4}$$

where:

$$\upsilon_+ = \frac{1}{2}\beta_e(g_D + g_A)B_0 + \sum_j\left(a_{Dj} + a_{Aj}\right)m_j \tag{S5a}$$

$$\upsilon_- = \frac{1}{2}\beta_e(g_D - g_A)B_0 + \sum_j\left(a_{Dj} - a_{Aj}\right)m_j \tag{S5b}$$

The angle $\theta$ follows the standard definition, the angle between the vector connecting the two spins and the laboratory magnetic field. More generally, the Hamiltonian matrix, eq. S4, is written in the laboratory frame as defined by the high-field approximation. It is readily apparent that the off-diagonal term, $\nu_-$, (i.e. the difference between the effective Zeeman energies of the two ions, also including the contribution from hyperfine couplings) will mix the $|T_0\rangle$ and $|S\rangle$ states and upon diagonalization yields the eigenstates:

$$\begin{aligned} &|1\rangle = |T_{+1}\rangle \\ &|2\rangle \equiv |\Phi_A\rangle = \cos\phi\,|S\rangle + \sin\phi\,|T_0\rangle \\ &|3\rangle \equiv |\Phi_B\rangle = -\sin\phi\,|S\rangle + \cos\phi\,|T_0\rangle \\ &|4\rangle = |T_{-1}\rangle \end{aligned} \tag{S6}$$

when

$$\phi = \frac{1}{2}\tan^{-1}\frac{2\upsilon_-}{D/2-J} \tag{S7}$$

And the eigenvalues:

$$
\begin{aligned}
E_1 &= \nu_+ + \frac{(J+D)}{4} \\
E_2 &= -\frac{(J+D)}{4} + \sqrt{\left(\frac{J}{2} - \frac{D}{4}\right)^2 + \nu_-^2} \\
E_3 &= -\frac{(J+D)}{4} - \sqrt{\left(\frac{J}{2} - \frac{D}{4}\right)^2 + \nu_-^2} \\
E_4 &= -\nu_+ + \frac{(J+D)}{4}
\end{aligned} \tag{S8}
$$

The spin Hamiltonian and its resulting eigenvalues and eigenvectors are independent of any molecular chirality and only depend on the magnetic properties of the involved spins.

The steady-state EPR spectrum (after coherences are lost) in field or frequency space is composed of peaks at positions determined by differences in the eigenvalues of the spin Hamiltonian, eqns. S8, with intensities determined by the transition rate between the eigenstates $|i\rangle$ and $|j\rangle$ and their population differences:

$$I_{ij} \propto \left|\langle i|\hat{S}_y|j\rangle\right|^2 (p_i - p_j) \tag{S9}$$

where $p_i$ and $p_j$ are the populations of the two eigenstates which correspond to the diagonal elements of the density matrix in the eigenbasis and depend upon on the initial state, i.e. $p_i = |\langle i|\psi_0\rangle|^2$. $\langle i|\hat{S}_y|j\rangle$ is the magnetic dipole matrix element of the examined transition along the direction of the microwave field, which is orthogonal to the static field.

Conversely, transition frequencies (and hence peak positions) only depend on the Hamiltonian eigenvalues and not on the specific initial state.

The four transitions of the SCRP (starting from a singlet precursor) appear at their usual positions. All four transitions are of the same intensity (see definition in Eq. S9) such that:

$$I_{12} = -I_{24} = I_{13} = -I_{34} = \frac{\cos^2\phi \sin^2\phi}{2} = \frac{\upsilon_-^2}{2(D/2-J)^2 + 8\upsilon_-^2} \tag{S10}$$

*SCRP with the CISS Effect*

The state for a radical pair following electron transfer through a chiral bridge can be written as

$$|\psi_0\rangle = \cos\chi\,|S\rangle + \sin\chi\,|T_0(\boldsymbol{n})\rangle \tag{S11}$$

This state was first proposed by Luo and Hore,(*46*) where the angle $\chi$ is related to the chirality and quantifies the effect of CISS in the reaction. The states $|S\rangle$ and $|T_0(\boldsymbol{n})\rangle$ are the radical pair singlet and the radical pair triplet states, respectively; in particular, the triplet state is defined with respect to the axis $\boldsymbol{n}$ of charge separation, which is also the axis of chirality, the CISS quantization axis, and the long molecular axis. In the limiting cases: $\chi = 0$ (no CISS), $|\psi_0\rangle$ is a pure singlet state; at the other extreme $\chi = \pm\frac{\pi}{4}$, so that $|\psi_0\rangle = |\uparrow\downarrow(\boldsymbol{n})\rangle$ or $|\downarrow\uparrow(\boldsymbol{n})\rangle$, i.e. it shows maximum local spin polarization (100% CISS); and when $\chi = \pm\frac{\pi}{2}$, $|\psi_0\rangle = |T_0(\boldsymbol{n})\rangle$ a pure triplet state. The ± refers to the two different enantiomers.

In the basis of the spin Hamiltonian, eqn. S4, the triplet state is defined in the high-field limit and polarized along the laboratory magnetic field. When working in orientations or distributions of orientations where the CISS axis is not aligned with the laboratory magnetic field, it becomes necessary to rotate the state as defined in the molecular frame to the laboratory frame:

$$e^{-i\hat{S}_y\theta}|\psi_0\rangle = |\psi_0(\theta)\rangle = \cos\chi\,|S\rangle + \sin\chi\left[\cos\theta\,|T_0\rangle + \frac{\sin\theta}{\sqrt{2}}(|T_{-1}\rangle - |T_{+1}\rangle)\right] \quad \text{(S12)}$$

where the singlet component is rotationally invariant, while $|T_0(\boldsymbol{n})\rangle \equiv |T_0(\theta)\rangle = \cos\theta\,|T_0\rangle + \frac{\sin\theta}{\sqrt{2}}(|T_{-1}\rangle - |T_{+1}\rangle)$. Similar to eqn. S2, $\theta$ is defined as the angle between the CISS axis and the magnetic field. In the $D^{\bullet+}$-B$\chi$-$A^{\bullet-}$ molecules studied here, the CISS axis and the dipolar axis are nearly collinear so $\theta$ in eqns. S2 and S12 refers to the same measure. When $\theta \neq 0$, sizable components on $|T_{-1}\rangle$ and $|T_{+1}\rangle$ are introduced (see Fig. 3 of the main text).

Let us now focus on the two limiting cases examined in this work, i.e. $\theta = 0°$ and $\theta = 90°$. The resulting intensity for $I_{13}$ and $I_{34}$ at $\theta = 90°$ is given by:

$$I_{13} = -I_{34} = \frac{1}{4}\cos^2\phi\,(\cos 2\chi - \cos^2\chi\cos 2\phi) \quad \text{(S13)}$$

Showing a dependence on the composition of the eigenstates ($\phi$) and on the amount of CISS ($\chi$). This can be separated into a singlet ($I_S$) and a 100% CISS ($I_{CISS}$) contributions, i.e.

$$\begin{aligned} I_{13} = -I_{34} &= \frac{1}{2}(1 - 2\sin^2\chi)\cos^2\phi\sin^2\phi - \frac{1}{4}\sin^2\chi\cos^2\phi\cos 2\phi \\ &= (1-p)\, I_S + p\, I_{CISS} \end{aligned} \tag{S14}$$

where $p = 2\sin^2\chi$ represents the CISS efficiency and

$$I_S = \frac{1}{2}\cos^2\phi\sin^2\phi \tag{S15}$$

$$I_{CISS} = -\frac{1}{8}\cos 2\phi\cos^2\phi \tag{S16}$$

Note that for this orientation $I_S$ and $I_{CISS}$ have opposite signs and that $I_{CISS}$ is the same that would be obtained starting from a fully polarized initial state ($\chi = \pm\pi/4$ ).

The calculation is perfectly analogous for $I_{12} = -I_{24}$, so that

$$I_{12} = -I_{24} = \frac{1}{4}\sin^2\phi\ \ (2\cos^2\phi\cos^2\chi - \sin^2\chi) \tag{S17}$$

This can also be decomposed into singlet and 100% CISS contributions, with the same sign. Remarkably, in the present case $I_{12}$ and $I_{24}$ are much weaker and occur in the central part of the spectrum. Hence, we focus on the main peaks $I_{34}$ and $I_{13}$ in the following. Similar expressions can also be derived for $\theta = 0$. A summary of the different singlet and CISS contributions for each orientation is reported in Table S5.

In Table S5, blue (orange) indicate emissive (absorptive) signals, fixed by the sign of *D*. Note that opposite enantiomers are indistinguishable when $\theta = 90°$, while they give different intensities when $\theta = 0°$, which are therefore averaged in the Table S5.

For these specific directions the same results could be obtained using the initial mixed state for electron transfer through a chiral bridge proposed by Chiesa et al.(*47*) or the coherent superposition

$$|\psi_0'\rangle = \cos\chi\,|S\rangle + i\sin\chi\,|T_0(\boldsymbol{n})\rangle \tag{S18}$$

derived by Fay.(*45*) Eq. S18 does not yield any single-spin polarization, but was expanded later by Fay(*44*) to account for spin-spin exchange interactions during multi-step electron transfer processes, leading to polarization effects similar to those postulated by Luo and Hore(*46*) (a mixture of states S18 and S11 with the same angle $\chi$). Since the three models are equivalent for describing the TREPR spectra, we mainly focus on Eq. S11.(*46*)

*TREPR Spectral Linewidths*

To understand how the entanglement of the eigenstates affects the linewidths for different initial states, let us focus again on $\theta = 90°$ and transition $I_{13}$ between $|\Phi_B\rangle$ and $|\mathrm{T}_{+1}\rangle$, which occurs on the low field side of the spectrum. Analogous reasoning holds for transition $I_{34}$ on the high field side of the spectrum. The intensities of the transitions for singlet and 100% CISS initial states are given by Eqs. S15 and S16 above. Their dependence on the mixing angle $\phi$ is very different, as shown in Fig. S10. In order to better compare the linewidths of the two curves, here we report the absolute value of $I_\mathrm{S}$ and $I_\mathrm{CISS}$ (Fig. S10A), along with its derivative (Fig. S10B).

As discussed in the main text, $\phi$ can be varied by considering different configurations of the nuclear spins coupled to the SCRP or by distributions of the Hamiltonian parameters. In the present calculations, we included a 14% strain in the donor-acceptor distance, leading to a distribution of dipole-dipole interactions. This, in turn, corresponds to resonances occurring at different values of $B_0$. As a result, we can have regions in the tails of a specific peak, which correspond to more entangled states, i.e. such that $\phi$ is closer to 0 compared to the center of the peak. In this case, moving from the center of the peak to the tail corresponds to moving toward $\phi = 0$ in Fig. S10B, where $d|I_\mathrm{CISS}|/d\phi$ is positive and $d|I_\mathrm{S}|/d\phi$ is negative. Such a situation, in which $I_\mathrm{CISS}$ increases while $I_\mathrm{S}$ decreases from the center to the tail of the peak leads to a larger broadening of $I_\mathrm{CISS}$ compared to $I_\mathrm{S}$, thus exactly producing CISS features in the tail when $I_\mathrm{CISS}$ and $I_\mathrm{S}$ are combined.

*TREPR Spectral Simulations*

The time dependence of the TREPR spectra is simulated by numerically integrating the stochastic Liouville equation for the system density matrix

$$\frac{d\rho_r(\Omega,t)}{dt} = -i\left[\frac{1}{\hbar}\,\tilde{H}(\Omega) + i\tilde{R} + i\tilde{K}\right]\rho_r(\Omega,t) \qquad \text{(S19)}$$

where $\rho_r(\Omega,t)$ is the density matrix of the system in the rotating frame, $\tilde{H}(\Omega)$, $\tilde{R}$ and $\tilde{K}$ are super-operators associated to the coherent Hamiltonian evolution, to dephasing/relaxation and recombination effects, respectively. Both relaxation and recombination are negligible in the time range of interest and we therefore omit them in the following. Conversely, off-diagonal elements of $\rho_r$ are affected by pure dephasing, parametrized by transition-dependent $T_2$.

Note that in the high-field limit, the non-secular terms in the Hamiltonian are also negligible and the rotating-frame Hamiltonian (including the microwave field) becomes time independent. Hence, Eq. S19 can be directly integrated, leading to

$$\rho_r(\Omega,t) = \exp\left[-\frac{i}{\hbar}\,\tilde{H}(\Omega) + \tilde{R} + \tilde{K}\right] t\,\rho_r(\Omega,0) \qquad \text{(S20)}$$

from which the spectrum $\langle S_y(B_0,t)\rangle = Tr[\rho_r(\Omega,t)S_y]$ is calculated.(*47*)

Apart from the rise of the signal at short times, the resulting spectra are equivalent to those obtained from linear-response theory in the stationary limit, Eq. S9, using, for instance, the Easyspin package for spin-polarized initial states.(*54, 55*)

The time and field dependencies of the TREPR spectra simulated for the protonated compound using the minimal model described in the main text and simulated for the deuterated case including a distribution of hyperfine fields generated by the four $^{1}H$ and two $^{14}N$ nuclear spins of $NDI^{\bullet-}$ are reported in Figs. S11 and S12. The experimental spectra are very well reproduced, both for the main peaks and for the characteristic CISS features at the two wings. The latter show practically the same time dependence as the main peaks, indicating that they are not transient but resonant peaks (also present in the steady-state spectra). The full distributions of hyperfine fields, *g*-tensors and dipolar strain included in this model are slightly different from those reported in the main text, but these small differences do not affect the origin of CISS features.

<u>Electron Transfer Rates and Energetics</u>

The free energies of reaction for (*R* or *S*) **1-*h*$_9$** (**-*d*$_9$**) and **2-*h*$_9$** (**-*d*$_9$**) were determined from the modified Weller equation(*58*) (Eq. S21) using the one-electron oxidation potential $E^{+}_{1/2}$ of PXX (0.81 V vs. SCE(*31*), and the one-electron reduction potentials $E^{-}_{1/2}$ of NMI, $NMI_2$, and NDI (-1.4 V, (*53*) -1.5 V(*59*) and -0.48 V (*53*) vs SCE, respectively), the estimated dielectric constant of PrCN at 105 K ($\varepsilon = 2.3$),(*60*) and the center-to-center radical ion pair distances given in Table S1 determined as noted. The ionic radii are estimated as half of the donor-acceptor distance and $e_0$ is the charge of the electron.

$$\Delta G_{DA} = E^{+}_{1/2} - E^{-}_{1/2} + \frac{e_0^2}{\varepsilon}\left(\frac{1}{2r_D} + \frac{1}{2r_A} - \frac{1}{r_{DA}}\right) \qquad \text{(S21)}$$

The values of the electronic coupling matrix elements for the electron transfer reaction were determined using the semi-classical expression for the electron transfer rate $k_{ET}$ given in Eq. S22, (*60*) where $\Delta G_{DA}$ and $k_{ET}$ are the free energies of reaction and measured rate constants given in Table S1. The values of $k_{ET}$ are obtained from the data presented in Figs. S5 and S6. In addition, the total internal reorganization for the one-electron oxidation of PXX and reduction of NMI, $NMI_2$ or NDI ($\lambda_i$) is estimated as 0.5 eV, while the solvent reorganization energy of PrCN at 105 K ($\lambda_S$) is estimated from Eq. S23(*60*) and we make the usual assumption of one dominant vibration ($\omega$ = 1500 $cm^{-1}$) coupled to the reaction coordinate of the electron transfer reaction.

$$k_{ET} = \frac{2\pi}{\hbar} V_{DA}^2 \frac{1}{(4\pi\lambda_s k_B T)^{1/2}} \sum_{m=1}^{\infty} \frac{s^m e^{-s}}{m!} \cdot \exp\left[-\frac{(\lambda_s + \Delta G_{DA} + m\omega)^2}{4\lambda_s k_B T}\right] \qquad \text{(S22)}$$

where $s = \lambda_i/\omega$

$$\lambda_s = e_0^2 \left(\frac{1}{2r_D} + \frac{1}{2r_A} - \frac{1}{r_{DA}}\right)\left(\frac{1}{n^2} - \frac{1}{\varepsilon}\right) \qquad \text{(S23)}$$

where *n* is the refractive index of the solvent and the remaining values are defined above.

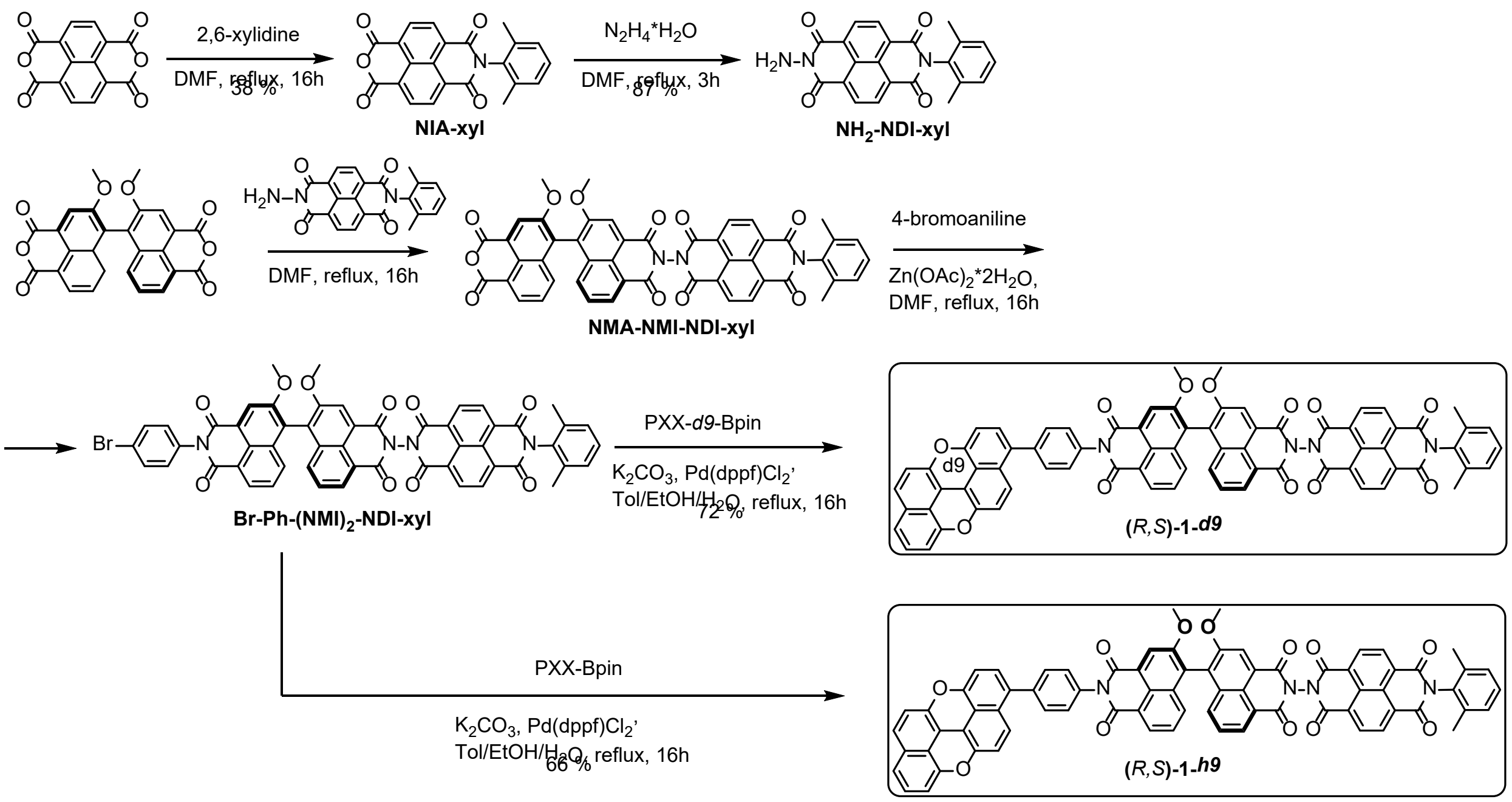

**Fig. S1. Synthetic scheme** for (*R* and *S*)-**1-*h*₉** and (*R* and *S*)-**1-*d*₉**

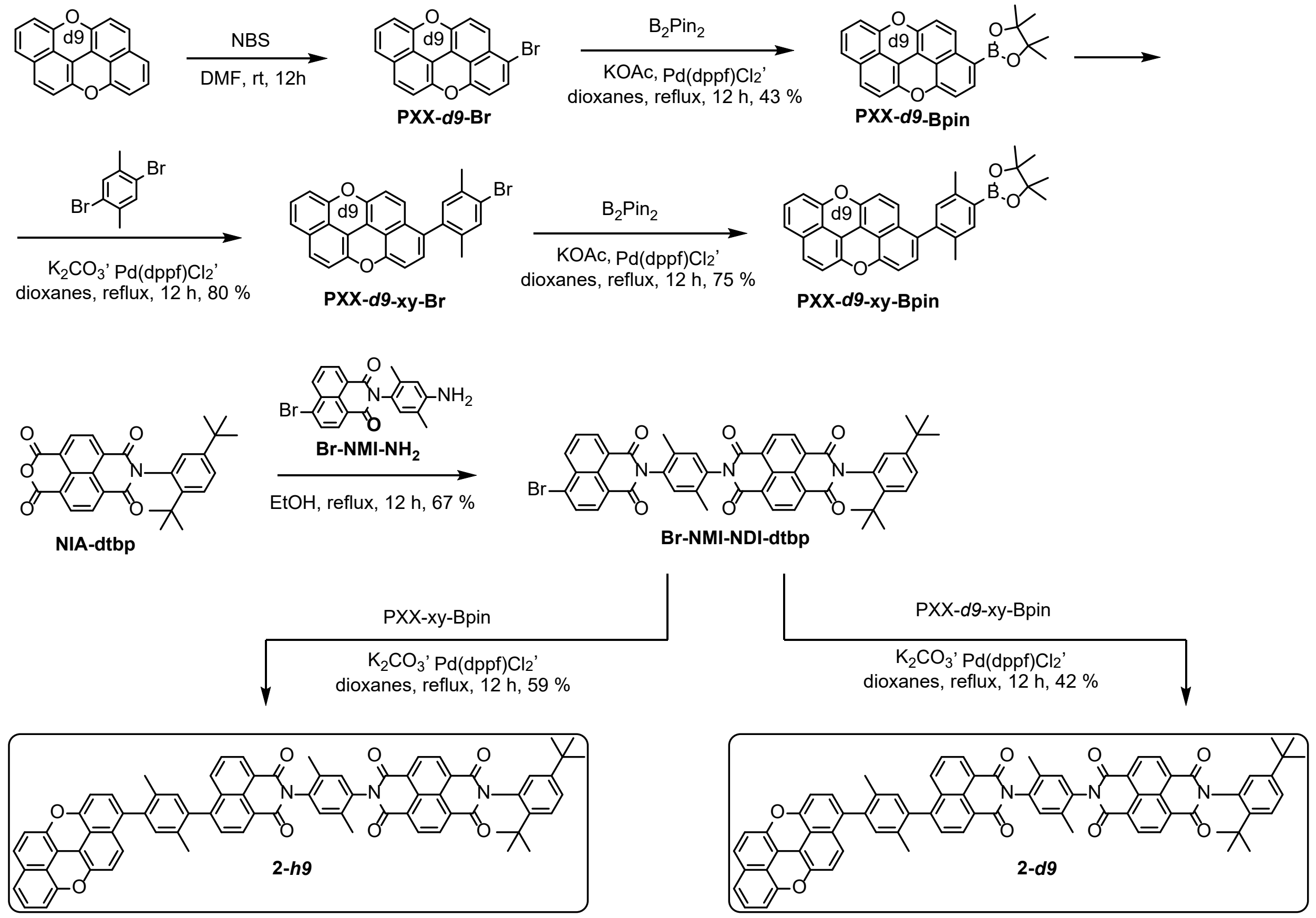

**Fig. S2. Synthetic scheme** for **2-*h*9** and **2-*d*9**

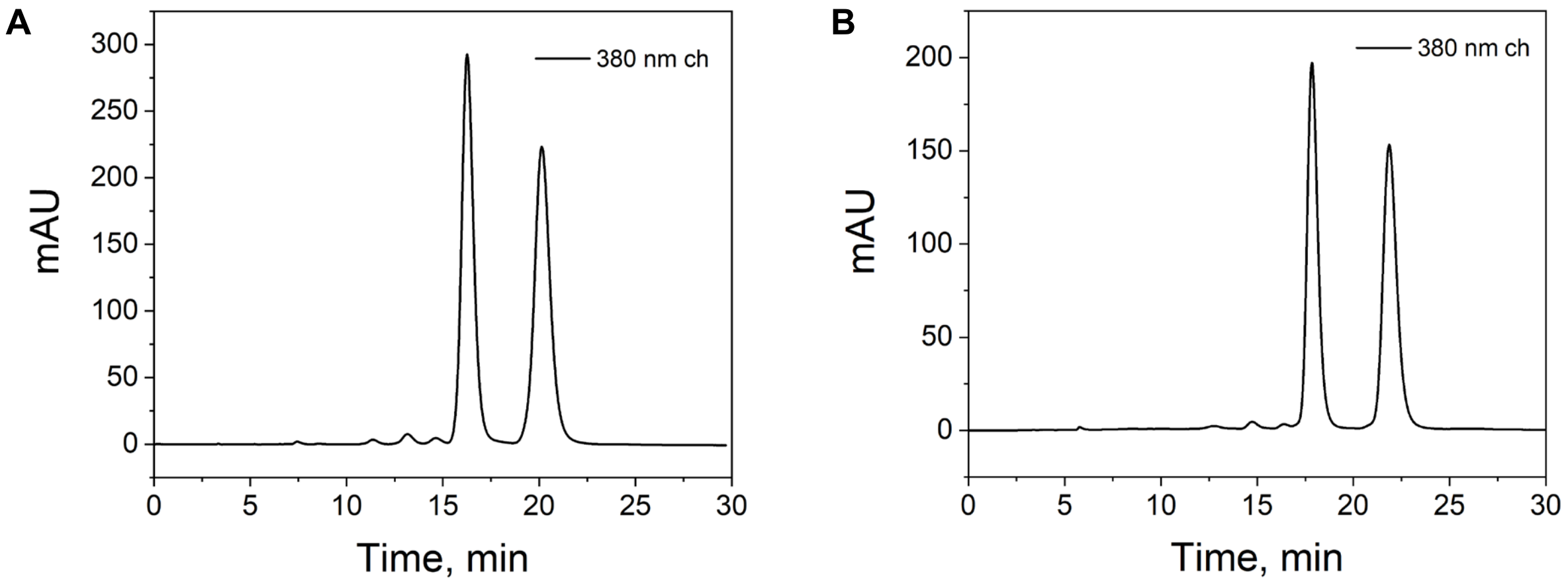

**Fig. S3. HPLC chromatogram** showing the chiral resolution of (**A**) (*R* and *S*)-**1-*h*$_9$** and (**B**) (*R* and *S*)-**1-*d*$_9$** enantiomers on a ChiralPak IA-3 column (250 x 4.6 mm, 3 µm) monitored at 380 nm using a dichloromethane/hexanes/isopropanol mixture as an eluent (50/45/5 volume %) at 1 mL/min flow rate. Peaks with the shorter retention times are the (*S*) enantiomers.

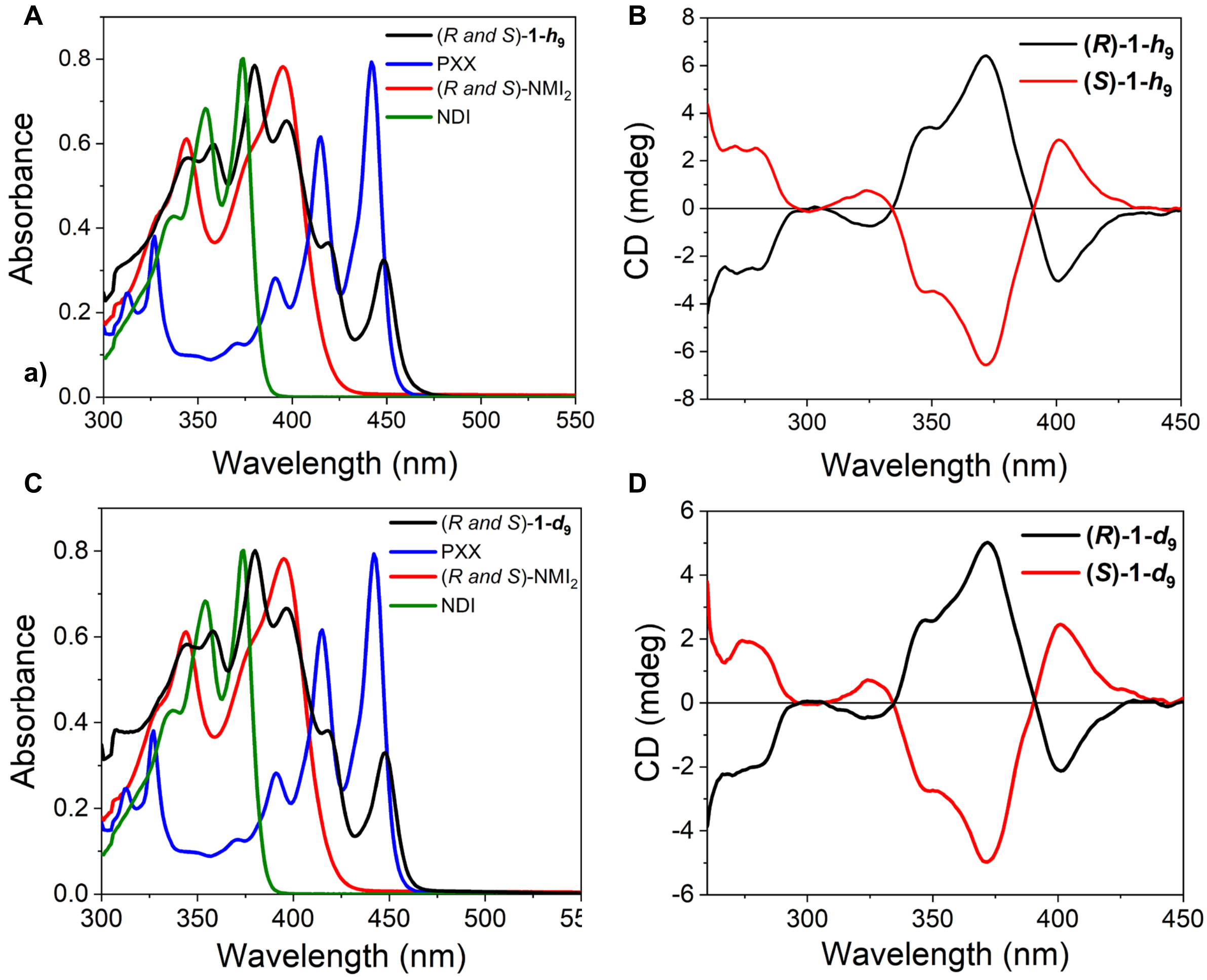

**Fig. S4. Optical spectra.** (**A**) UV-vis absorption of the racemic mixture of (*R* and *S*)-**1-*h*9,** PXX, (*R* and *S*)-NMI2 and NDI in toluene at 295 K. (**B**) CD spectra of (*R*)-**1-*h*9** and (*S*)-**1-*h*9**. (**C**) UV-vis absorption of the racemic mixture of (*R* and *S*)-**1-*d*9,** PXX, (*R* and *S*)-NMI2 and NDI in toluene at 295 K. (**D**) CD spectra of (*R*)-**1-*d*9** and (*S*)-**1-*d*9**.

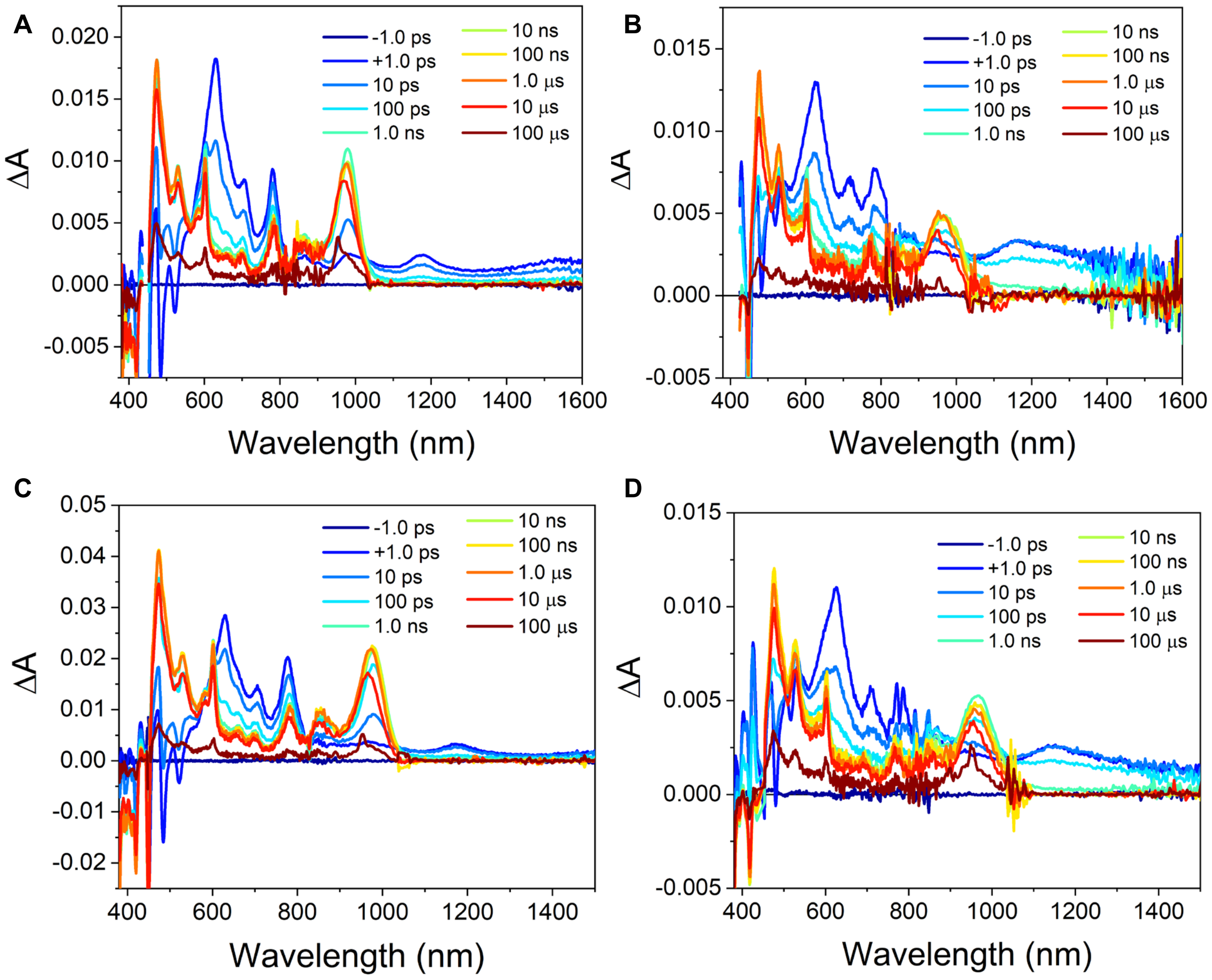

**Fig. S5. Transient absorption data** following 450 nm excitation of compounds in PrCN at 105 K at the indicated pump-probe delay times. (**A**) (*R* and *S*)-**1-*h*₉**, (**B**) **2-*h*₉**, (**C**) (*R* and *S*)-**1-*d*₉**, (**D**) **2-*d*₉**.

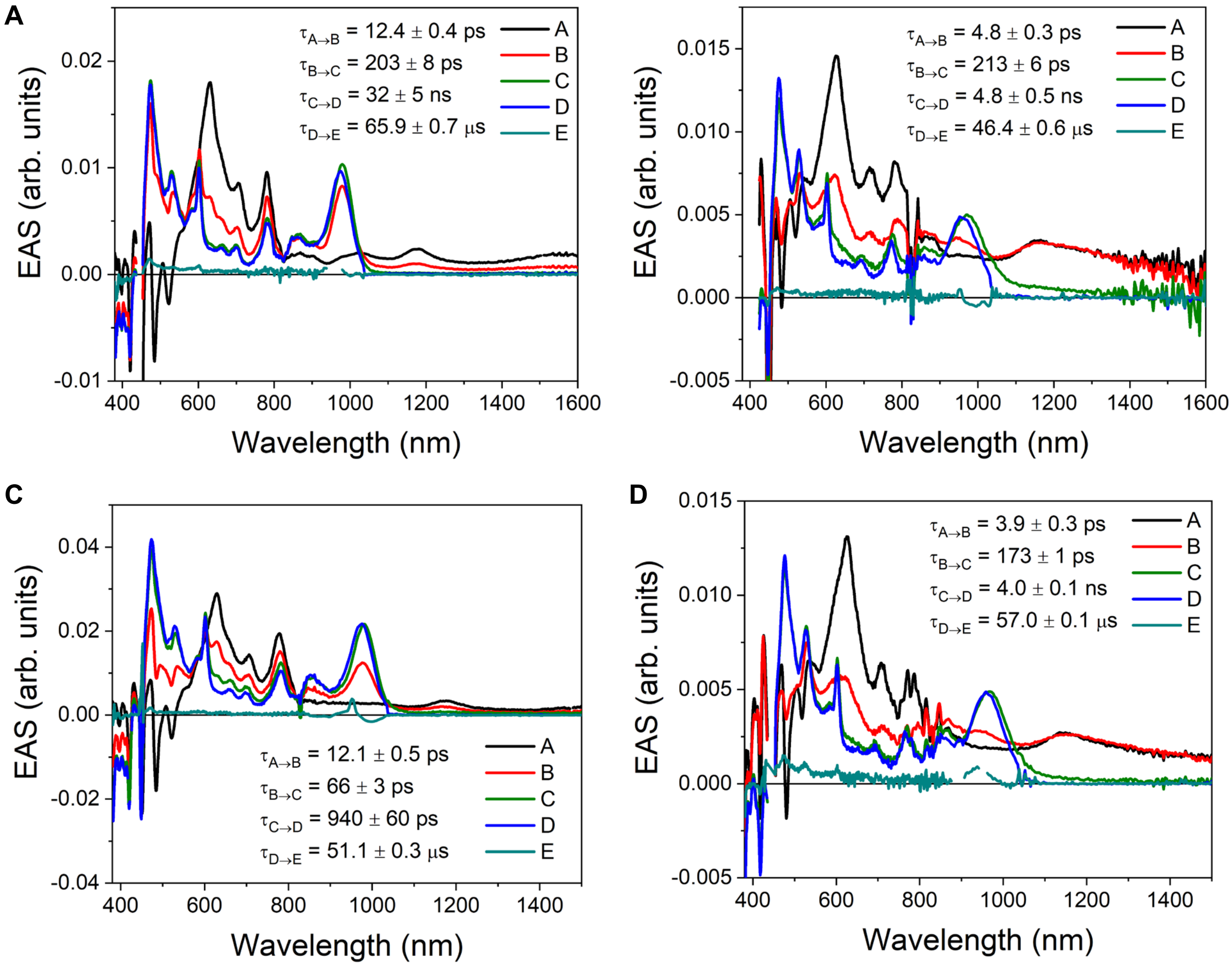

**Fig. S6. Evolution-associated spectra** obtained by globally fitting the femto- and nanosecond transient absorption data for (**A**) (*R* and *S*)-**1-*h*$_9$**, (**B**) **2-*h*$_9$**, (**C**) (*R* and *S*)-**1-*d*$_9$**, and (**D**) **2-*d*$_9$**, following 450 nm excitation in PrCN at 105 K.

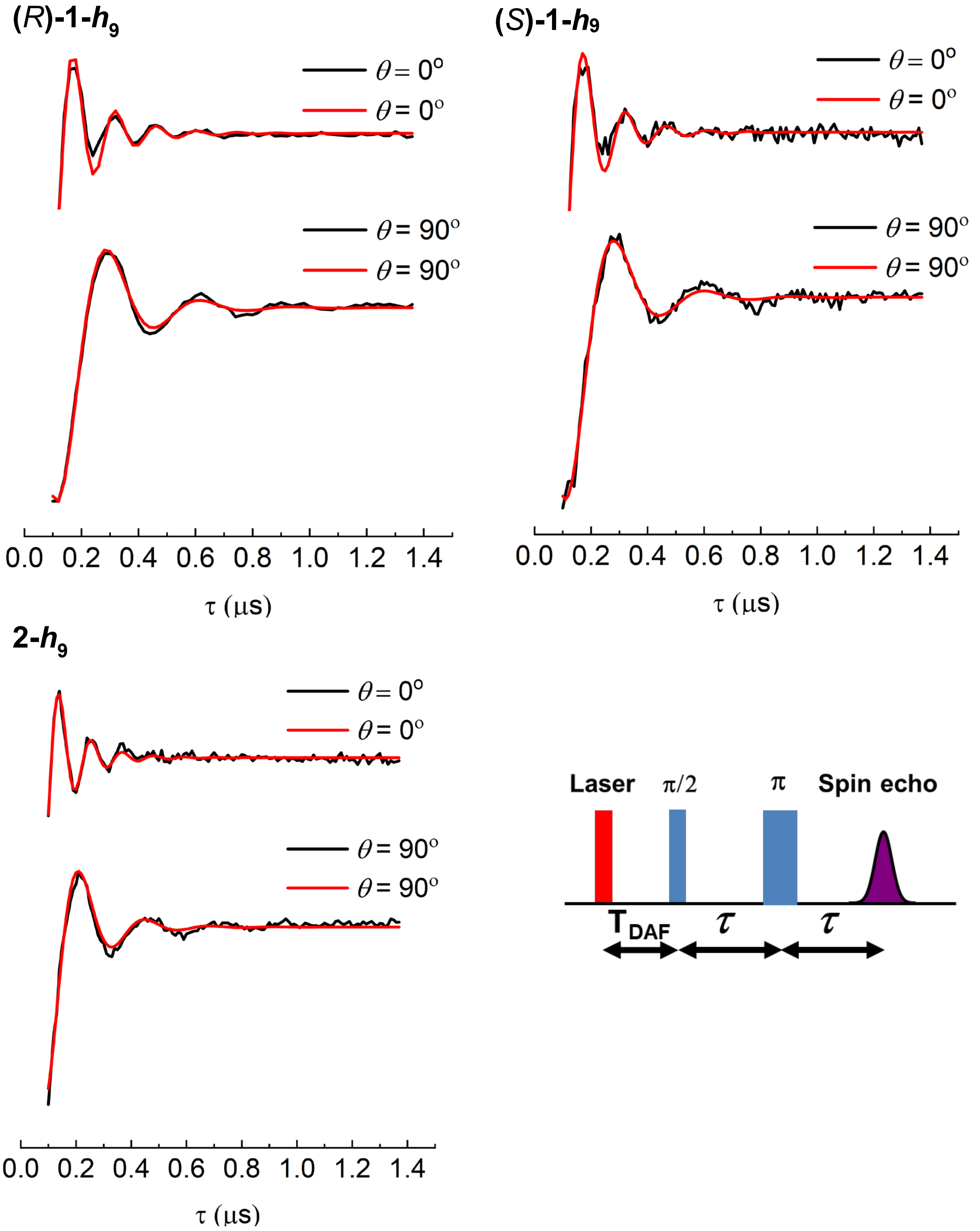

**Fig. S7. OOP-ESEEM data and fits** at $\theta = 0°$ and $\theta = 90°$ for the indicated molecules oriented in 5CB. Lower right: Hahn echo pulse sequence for OOP-ESEEM experiments, where $T_{DAF}$ is the delay after the laser flash.

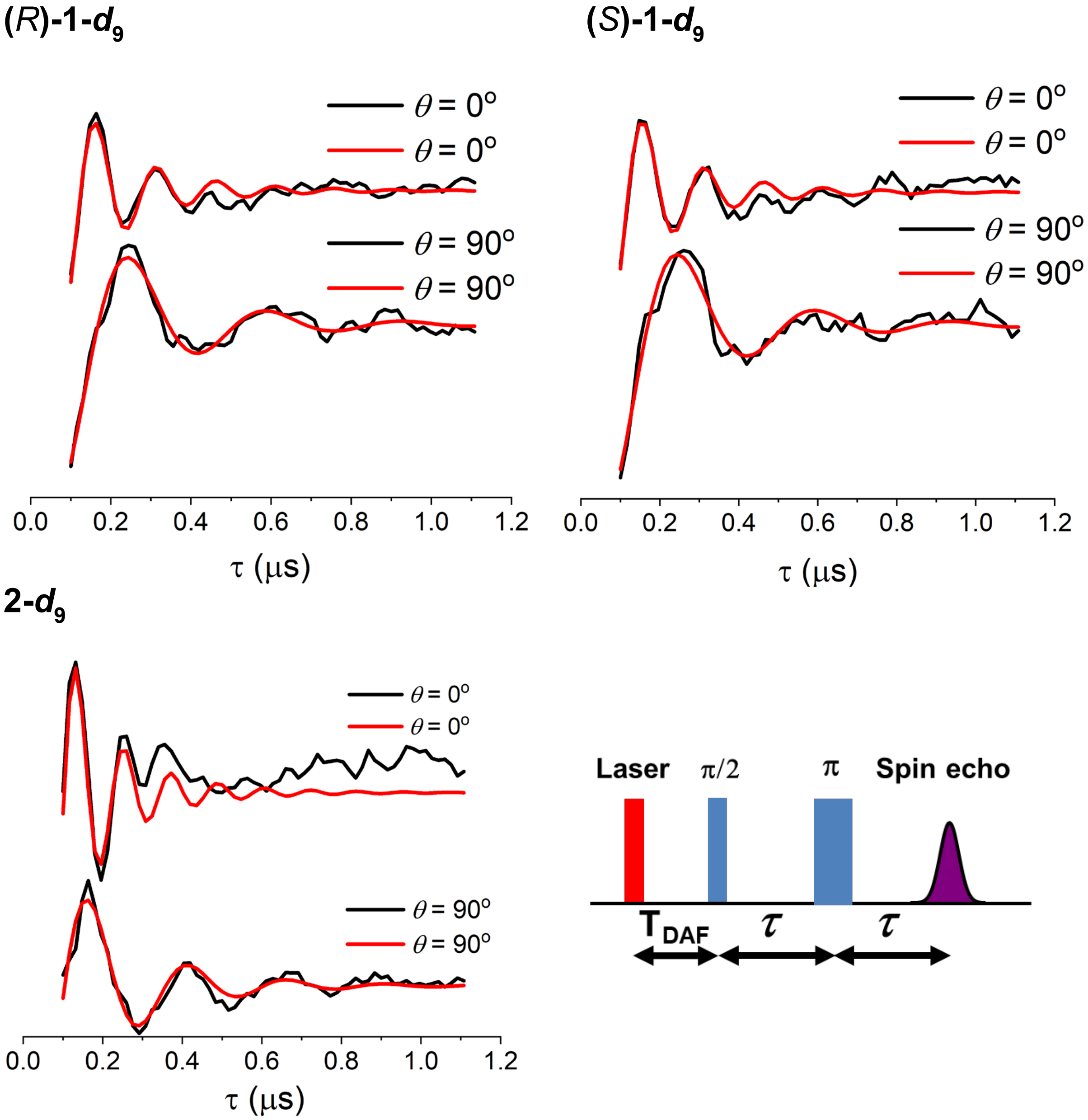

**Fig. S8. OOP-ESEEM data and fits** at $\theta = 0°$ and $\theta = 90°$ for the indicated molecules oriented in 5CB. Lower right: Hahn echo pulse sequence for OOP-ESEEM experiments, where $T_{DAF}$ is the delay after the laser flash.

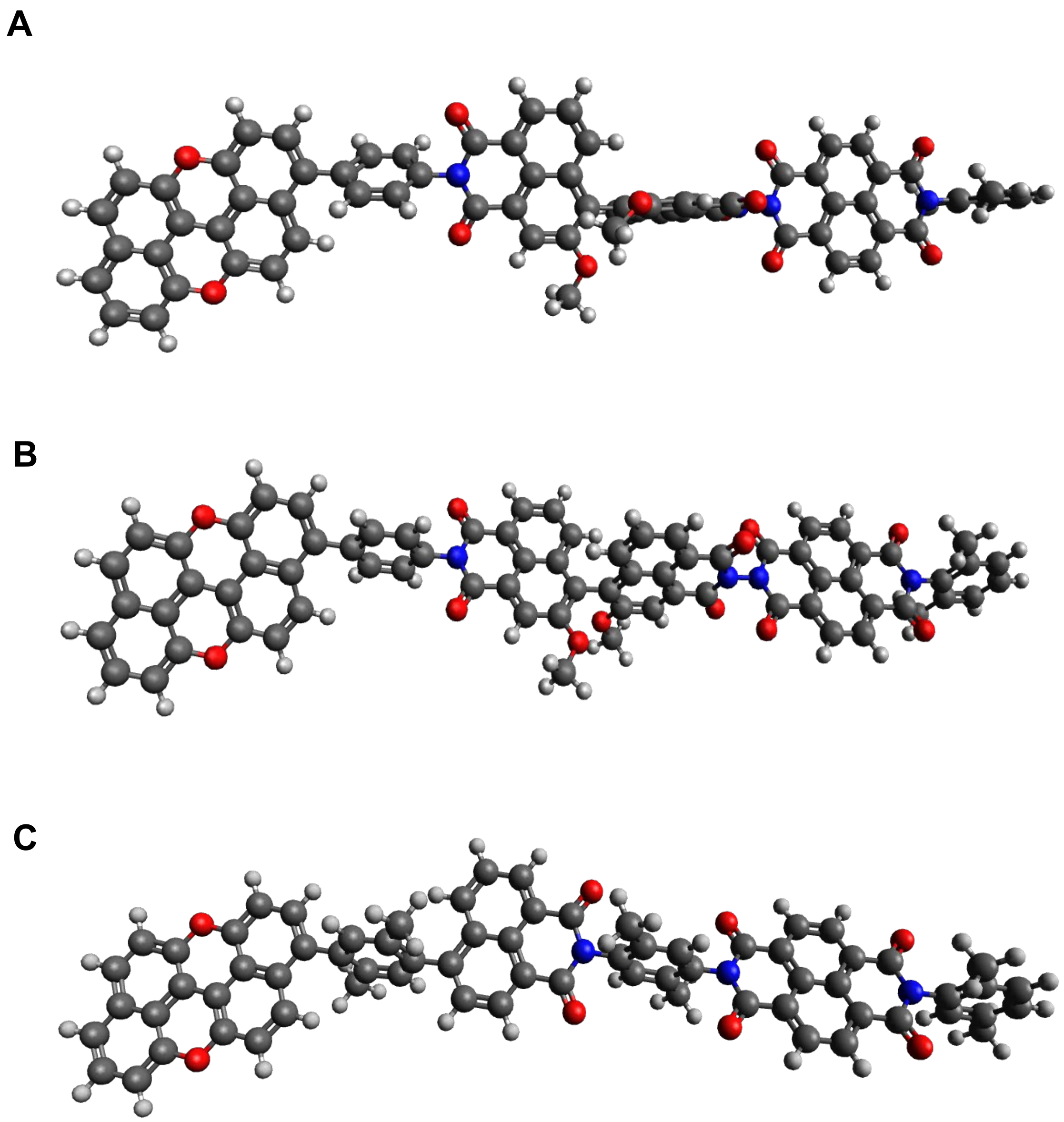

**Fig. S9. Geometry optimized structures.** (**A**) (*R*)-**1-*h*$_9$**, (**B**) (*S*)-**1-*h*$_9$**, and (**C**) **2-*h*$_9$** obtained using density functional theory (DFT) at the B3LYP/6-31G* level of theory in QChem (version 5.1).

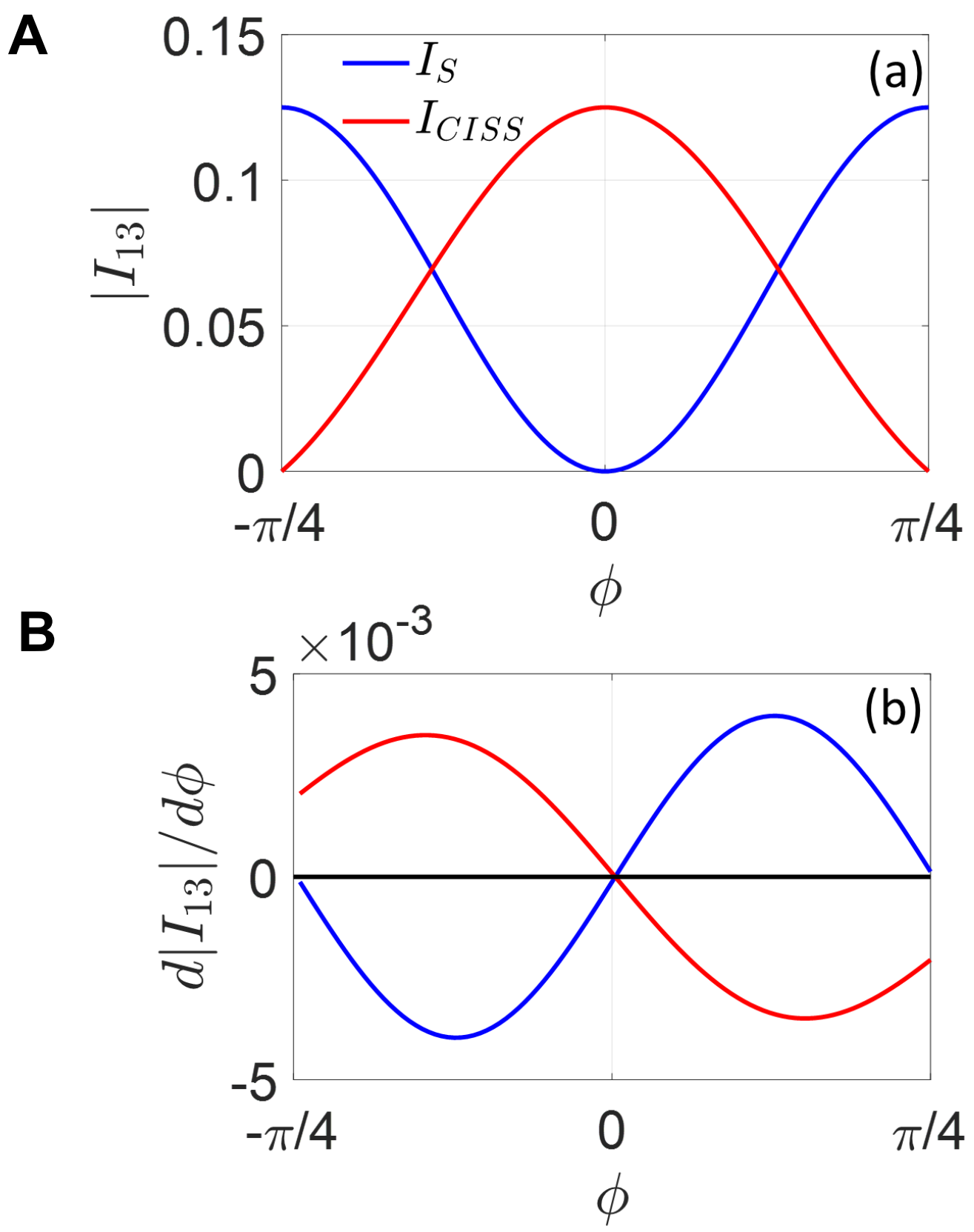

**Fig. S10. Transition intensity as a function of the CISS contribution.** (**A**) Dependence of the absolute value of the intensity of the $|\Phi_B\rangle \leftrightarrow |T_{+1}\rangle$ (1-3) transition on the composition of the eigenstates, expressed by the angle $\phi$, for an initial singlet (blue line) or fully polarized (100% CISS, red curve) state. (**B**) Same for its derivative.

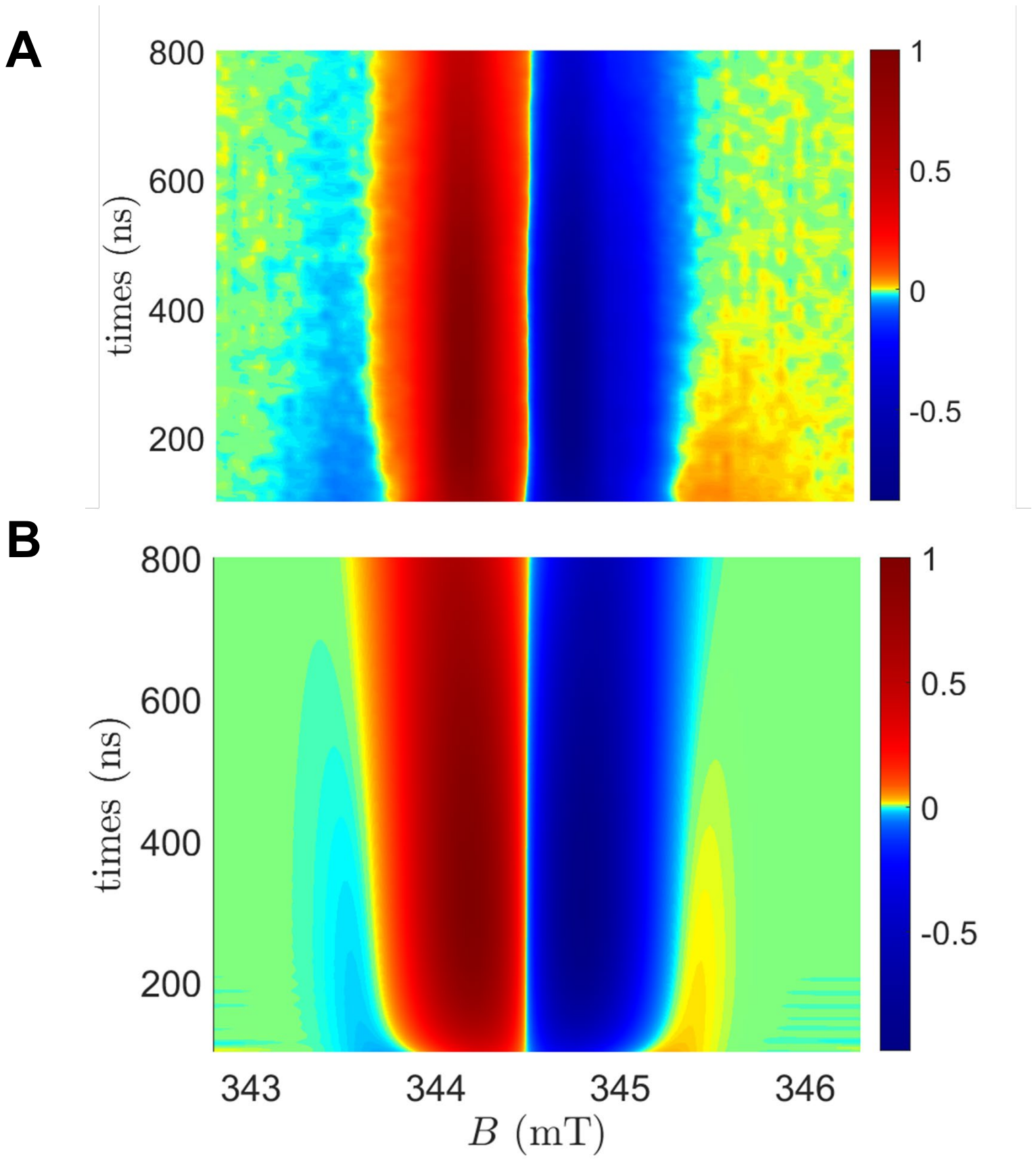

**Fig. S11. Time and field dependence of TREPR spectra.** Experimental (**A**) and simulated (**B**) time and field dependence of TREPR spectra of (*R* and *S*)-**1-*h*$_9$** with the long axis of each molecule perpendicular to the applied magnetic field direction. The time dependence of the lateral CISS features is enhanced using a non-linear color-scale.

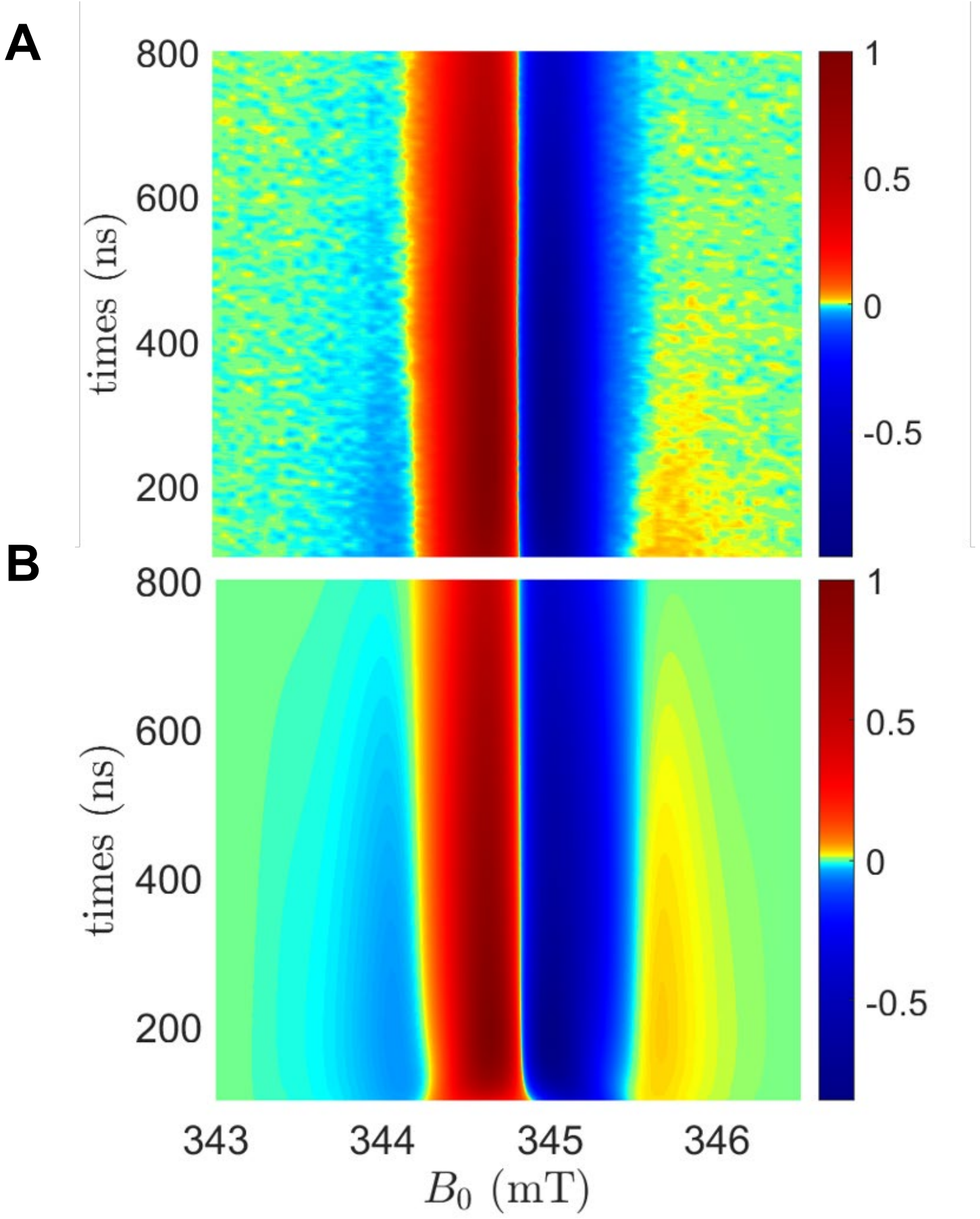

**Figure S12. Time and field dependence of TREPR spectra.** Experimental (**A**) and simulated (**B**) time and field dependence of TREPR spectra of (*R* and *S*)-**1-*d*$_9$** with the long axis of each molecule perpendicular to the applied magnetic field direction. The time dependence of the lateral CISS features is enhanced using a non-linear color-scale. The effect of hyperfine interactions is included as a secular term in the Hamiltonian, with four $^1$H ($I = 1/2$) and two $^{14}$N ($I = 1$) nuclear spins coupled to $NDI^-$ with isotropic $a_H = 4$ MHz and $a_N = 2$ MHz. For each configuration of the nuclear spins, we consider a distribution of hyperfine fields with a gaussian broadening of 1 MHz to account for possible strain and anisotropy of the hyperfine couplings. A distribution of donor-acceptor distances with standard deviation 0.5 nm was also included. Slightly different values of the *g*-tensors compared to the minimal model were used $g_{NDI} = (2.0034, 2.0041, 2.0043)$, $g_{PXX} = (2.0032, 2.0046, 2.0046)$ and a CISS contribution of 47 %.

| Compound | Electron Transfer Reaction | $r_{DA}$ (nm) | $\Delta G_{DA}$ (eV) | $k_{ET}$ ($s^{-1}$) | $V_{DA}$ ($cm^{-1}$) |
|---|---|---|---|---|---|
| (*R* or *S*) **1-*h*₉**<br>(*R* or *S*) **1-*d*₉** | $^{1*}$PXX-NMI$_2$-NDI→<br>PXX$^{\bullet+}$-NMI$_2^{\bullet-}$-NDI | 1.46 [a] | -0.26 | 8.1 x $10^{10}$<br>8.3 x $10^{10}$ | 18<br>18 |
| (*R* or *S*) **1-*h*₉**<br>(*R* or *S*) **1-*d*₉** | PXX$^{\bullet+}$-NMI$_2^{\bullet-}$-NDI→<br>PXX$^{\bullet+}$-NMI$_2$-NDI$^{\bullet-}$ | 2.48 ± 0.01 [b] | -1.09 | 4.9 x $10^9$<br>1.5 x $10^{10}$ | 40<br>70 |
| **2-*h*₉**<br>**2-*d*₉** | $^{1*}$PXX-NMI-NDI→<br>PXX$^{\bullet+}$-NMI$^{\bullet-}$-NDI | 1.21 [a] | -0.21 | 2.1 x $10^{11}$<br>2.6 x $10^{11}$ | 74<br>83 |
| **2-*h*₉**<br>**2-*d*₉** | PXX$^{\bullet+}$-NMI$^{\bullet-}$-NDI→<br>PXX$^{\bullet+}$-NMI-NDI$^{\bullet-}$ | 2.28 ± 0.01 [b] | -1.12 | 4.7 x $10^9$<br>5.8 x $10^9$ | 18<br>20 |

**Table S1. Electron Transfer Data.** [a]Distances obtained from the DFT calculations given in Tables S3-S5. [b]Distances obtained from the OOP-ESEEM data given in the main text.

| Compound | *J* (MHz) | $r_{DA}$ (nm) |
|---|---|---|
| (*S*)-**1-*h*$_9$** | -0.21 ± 0.09 | 2.48 ± 0.01 |
| (*R*)-**1-*h*$_9$** | -0.28 ± 0.11 | 2.48 ± 0.01 |
| **2-*h*$_9$** | -0.14 ± 0.10 | 2.28 ± 0.01 |
| (*S*)-**1-*d*$_9$** | -0.31 ± 0.10 | 2.53 ± 0.01 |
| (*R*)-**1-*d*$_9$** | -0.18 ± 0.10 | 2.51 ± 0.01 |
| **2-*d*$_9$** | -0.25 ± 0.10 | 2.29 ± 0.01 |

**Table S2. OOP-ESEEM fit parameters**

(*R*)-**1*h*₉**

Coordinates (Angstroms)

| ATOM | | X | Y | Z |
|---|---|---|---|---|
| 1 | H | 16.3358942944 | -4.3281626064 | -2.2877779347 |
| 2 | H | 18.8019623042 | -4.1693433187 | -1.9479313717 |
| 3 | C | 16.7611601820 | -3.5145988373 | -1.7095809383 |
| 4 | C | 18.1587795359 | -3.4119561690 | -1.5083550495 |
| 5 | O | 14.5660393913 | -2.6739519154 | -1.3610844712 |
| 6 | H | 11.9554419289 | -2.7139744068 | -1.5808736086 |
| 7 | C | 15.9315444646 | -2.5615966536 | -1.1581652039 |
| 8 | H | 6.7328227085 | 1.1956699080 | 2.5948679097 |
| 9 | C | 12.3283682942 | -1.8685093007 | -1.0113134439 |
| 10 | C | 18.7163833194 | -2.3845926757 | -0.7749051188 |
| 11 | H | 9.1962546309 | 1.1050068920 | 2.8402936355 |
| 12 | C | 13.7165528883 | -1.7324192465 | -0.8150725568 |
| 13 | H | 19.7921984943 | -2.3275352957 | -0.6334012584 |
| 14 | C | 16.4771104203 | -1.4935031639 | -0.3987795457 |
| 15 | C | 11.4606382462 | -0.9370578779 | -0.4711447059 |
| 16 | H | -16.6065064634 | -2.6405937900 | 2.4174190066 |
| 17 | C | 7.3691531102 | 1.2034762054 | 1.7159872678 |
| 18 | C | -1.8269945346 | -0.5152290510 | 0.2083806265 |
| 19 | C | 17.8825534162 | -1.3907657745 | -0.1961869238 |
| 20 | C | -3.2551505378 | -0.4921919533 | 0.1830938352 |
| 21 | H | 10.3942787638 | -1.0680207988 | -0.6182288983 |
| 22 | C | 8.7541789852 | 1.1581587223 | 1.8496320274 |
| 23 | C | 14.1997169495 | -0.6752613331 | -0.0733415615 |
| 24 | O | -6.0321160099 | 1.1785465171 | -1.5033436950 |
| 25 | H | 1.6450232955 | -3.3923115486 | -1.3680027496 |
| 26 | C | -15.8988961538 | -2.0526615079 | 1.8260684184 |
| 27 | C | 15.6076544587 | -0.5298791633 | 0.1585942148 |
| 28 | H | -15.2557354228 | -1.5043821588 | 2.5250233216 |
| 29 | H | -15.2645627919 | -2.7546285296 | 1.2714685306 |
| 30 | C | 11.9189161302 | 0.1811046448 | 0.2893023948 |
| 31 | C | 18.3494238737 | -0.2892156097 | 0.5804127404 |
| 32 | C | 13.3273161708 | 0.2848801528 | 0.4886757962 |
| 33 | C | 2.5580440635 | -0.7477155995 | -0.9855300313 |
| 34 | C | 3.2308530614 | 0.3068324835 | -0.3978768148 |
| 35 | C | 4.7117015711 | 0.2278505312 | -0.2925212371 |
| 36 | C | 3.2246847375 | 2.5106941690 | 0.6817524741 |
| 37 | C | 2.5239245555 | 1.4266326195 | 0.0935108442 |
| 38 | C | 4.7040872449 | 2.4832567201 | 0.7968925806 |
| 39 | C | 1.1519314424 | -0.7219852159 | -1.0951769561 |
| 40 | C | 2.5343610706 | 3.6077804483 | 1.1614267250 |
| 41 | C | 1.1250477687 | -2.8658485725 | -2.1775722624 |
| 42 | O | 5.3427683922 | -0.7345995100 | -0.7007680978 |
| 43 | C | 1.0964757572 | 1.4573806069 | -0.0020593889 |
| 44 | C | 1.1273931851 | 3.6459799453 | 1.0751780471 |
| 45 | C | 0.4241377198 | 2.6007817471 | 0.5143995436 |
| 46 | C | 0.4108253560 | 0.3559605914 | -0.5997675805 |
| 47 | N | 5.3586783224 | 1.3246848202 | 0.3099582008 |
| 48 | O | 5.3426153314 | 3.3978093410 | 1.2893845788 |
| 49 | O | 0.4381285903 | -1.7149465830 | -1.6950858853 |
| 50 | C | 6.7970869904 | 1.2652523619 | 0.4464860312 |
| 51 | C | -1.0813041063 | 0.3342850776 | -0.6792930564 |
| 52 | H | 3.1449148758 | -1.5798425606 | -1.3545876724 |
| 53 | C | -3.9199877312 | 0.3852762671 | -0.7127643152 |
| 54 | C | -5.3988525559 | 0.4490882673 | -0.7649716890 |
| 55 | H | 3.0942621527 | 4.4251178485 | 1.6031474860 |
| 56 | H | 19.4179921686 | -0.1870363654 | 0.7497114800 |
| 57 | C | 16.0969213299 | 0.5186390063 | 0.9073452845 |
| 58 | C | 9.5925612131 | 1.1646275382 | 0.7234413754 |
| 59 | H | 0.3521351747 | -3.5129393114 | -2.5943785084 |
| 60 | H | 0.5916391273 | 4.5093864749 | 1.4583611731 |
| 61 | H | -18.5847437669 | -1.7471402971 | 1.5416830154 |
| 62 | H | -0.6578046941 | 2.6441050410 | 0.4623305454 |
| 63 | C | 11.0714018476 | 1.1706564492 | 0.8914466326 |
| 64 | C | 17.4853523455 | 0.6445700690 | 1.1210554482 |
| 65 | C | 13.8817253115 | 1.3359251312 | 1.2666464753 |
| 66 | C | -16.6350630992 | -1.1167235119 | 0.8979254592 |
| 67 | C | -18.0334175847 | -1.0858628676 | 0.8780366665 |
| 68 | O | 15.2466658104 | 1.4558545745 | 1.4589305555 |
| 69 | H | 1.8445339355 | -2.6047087247 | -2.9631668082 |
| 70 | C | 11.6596541329 | 2.1660526720 | 1.6594099405 |
| 71 | C | 7.6098222708 | 1.2742443630 | -0.6864641475 |
| 72 | C | -1.7786537054 | 1.1604887390 | -1.5592304434 |
| 73 | H | 17.8546567885 | 1.4771307987 | 1.7115359085 |
| 74 | C | 8.9942693199 | 1.2183667183 | -0.5466360004 |
| 75 | C | -3.1921607771 | 1.1886987654 | -1.5616667162 |
| 76 | C | 13.0536496921 | 2.2657897712 | 1.8551489924 |
| 77 | N | -6.0600321054 | -0.4226177944 | 0.1289480289 |
| 78 | C | -8.0720450190 | -1.2775784878 | -0.7953395931 |
| 79 | C | -8.0587675521 | 0.5224447933 | 0.9917241579 |
| 80 | O | -7.4190137550 | -2.0155396053 | -1.5056271643 |
| 81 | N | -7.4419081151 | -0.3896705828 | 0.1055825028 |
| 82 | O | -7.3947108199 | 1.2341969395 | 1.7182648899 |
| 83 | C | -9.5554276085 | -1.2266232375 | -0.7924381444 |
| 84 | C | -9.5425210188 | 0.5283720875 | 0.9540374904 |
| 85 | C | -10.2666136387 | -2.0587509152 | -1.6416549417 |
| 86 | C | -10.2396610820 | -0.3368914872 | 0.0735730623 |
| 87 | C | -10.2413279431 | 1.3857385175 | 1.7882821826 |
| 88 | C | -11.6733486674 | -2.0321755363 | -1.6542399872 |
| 89 | C | -11.6615519420 | -0.3124059291 | 0.0594812100 |
| 90 | C | -11.6481530585 | 1.4075665373 | 1.7729162819 |
| 91 | C | -12.3661936008 | -1.1733966689 | -0.8166332991 |
| 92 | C | -12.3534243734 | 0.5717515708 | 0.9225113922 |
| 93 | C | -13.8547414752 | -1.1637687658 | -0.8420941530 |
| 94 | C | -13.8417432800 | 0.6077046860 | 0.9240729942 |
| 95 | H | -9.7168837727 | -2.7294739324 | -2.2935061538 |
| 96 | O | -14.4982786817 | -1.8899765993 | -1.5810889678 |
| 97 | H | -9.6819744803 | 2.0367884286 | 2.4518067997 |
| 98 | N | -14.4962387750 | -0.2696743019 | 0.0372911183 |
| 99 | O | -14.4746688691 | 1.3535231217 | 1.6525687876 |
| 100 | C | -15.9494855099 | -0.2535365507 | 0.0320715017 |
| 101 | H | -12.2370578485 | -2.6812570916 | -2.3158569210 |
| 102 | H | 11.0205994208 | 2.9199799813 | 2.1104896529 |
| 103 | H | -12.2021268922 | 2.0753583226 | 2.4240462811 |
| 104 | C | -18.7183666280 | -0.2241967210 | 0.0245121917 |
| 105 | H | 13.4782569082 | 3.0670740983 | 2.4505982904 |
| 106 | H | 7.1606246412 | 1.3273192063 | -1.6726538926 |
| 107 | H | -19.8049880737 | -0.2126463569 | 0.0215495427 |
| 108 | H | 9.6213876730 | 1.2450920080 | -1.4330661484 |
| 109 | C | -5.4722302084 | -1.3233034767 | 1.0492664600 |
| 110 | C | -16.6116549564 | 0.6239506080 | -0.8372798088 |
| 111 | C | -18.0105375461 | 0.6227323822 | -0.8250945128 |
| 112 | O | -6.1716778923 | -2.0185844561 | 1.7609092303 |
| 113 | C | -15.8511161853 | 1.5451770774 | -1.7603172858 |
| 114 | H | -18.5437857318 | 1.2958553438 | -1.4916735287 |
| 115 | H | -15.2088283665 | 2.2361038363 | -1.2009031220 |
| 116 | H | -15.2123475188 | 0.9847053983 | -2.4535991740 |
| 117 | H | -16.5432251711 | 2.1452427082 | -2.3577702042 |
| 118 | C | -3.9925083722 | -1.3309211117 | 1.0605510605 |
| 119 | C | -3.3400425239 | -2.1681109004 | 1.9475446532 |
| 120 | H | -3.9317792663 | -2.7964165193 | 2.6048264370 |
| 121 | C | -1.9338814917 | -2.1837520132 | 1.9893229627 |
| 122 | H | -1.4254090223 | -2.8346269061 | 2.6945049820 |
| 123 | C | -1.1967994922 | -1.3781139013 | 1.1451306973 |
| 124 | H | -0.1140547656 | -1.3958543095 | 1.1976229668 |
| 125 | O | -1.1924185285 | 2.0348260952 | -2.4318562869 |
| 126 | C | -0.1820177022 | 1.5633367239 | -3.3307740400 |
| 127 | H | -0.2906188841 | 2.1543016481 | -4.2438657364 |
| 128 | H | -0.3248378460 | 0.5022055350 | -3.5609310127 |
| 129 | H | 0.8199678125 | 1.7168091366 | -2.9192293300 |
| 130 | H | -3.7019977033 | 1.8525417974 | -2.2509995110 |

**Table S3. Optimized cartesian coordinates** for (*R*)-**1**-***h*₉**.

**(*S*)-1-*h*₉**

Coordinates (Angstroms)

| ATOM | | X | Y | Z |
|---|---|---|---|---|
| 1 | C | 17.4475492751 | -1.7348908307 | 0.4157596817 |
| 2 | C | 18.3988117941 | -0.7584104561 | 0.6446031257 |
| 3 | C | 18.0515751565 | 0.6175297370 | 0.7872092214 |
| 4 | C | 16.6717480335 | 0.9529257366 | 0.6810054414 |
| 5 | C | 15.7123688121 | -0.0564053915 | 0.4446238925 |
| 6 | C | 16.0871814926 | -1.3764790487 | 0.3159447150 |
| 7 | C | 18.9788188713 | 1.6668239536 | 1.0264080357 |
| 8 | C | 18.5336977517 | 2.9670751178 | 1.1500370600 |
| 9 | C | 17.1617779316 | 3.3006831816 | 1.0455022680 |
| 10 | C | 16.2430600744 | 2.2993790511 | 0.8133484350 |
| 11 | C | 14.3346873559 | 0.3301272277 | 0.3455613577 |
| 12 | C | 13.3723245058 | -0.6791075337 | 0.1128919782 |
| 13 | C | 13.8101192750 | -2.0253690700 | 0.0031291594 |
| 14 | O | 15.1490722085 | -2.3638475260 | 0.0915262602 |
| 15 | O | 14.9034385863 | 2.6385697819 | 0.7165072235 |
| 16 | C | 13.9660633291 | 1.6513072205 | 0.4870169573 |
| 17 | C | 11.9893752169 | -0.3446129856 | 0.0093429096 |
| 18 | C | 11.0472652202 | -1.4037220693 | -0.2181172140 |
| 19 | C | 11.5240049830 | -2.7048787962 | -0.3006010101 |
| 20 | C | 12.8922482857 | -3.0329406261 | -0.1942869281 |
| 21 | C | 12.6049329114 | 2.0052299597 | 0.4114034295 |
| 22 | C | 11.6504307121 | 1.0312654018 | 0.1823731182 |
| 23 | C | 9.5863189140 | -1.1584891203 | -0.3634120439 |
| 24 | C | 9.0756958417 | -0.2777075887 | -1.3322511746 |
| 25 | C | 7.7023752498 | -0.1067355114 | -1.4936929360 |
| 26 | C | 6.8174365157 | -0.8170310671 | -0.6854696566 |
| 27 | C | 7.3014025771 | -1.6920909986 | 0.2844916947 |
| 28 | C | 8.6743732683 | -1.8584342995 | 0.4430025136 |
| 29 | N | 5.3887179256 | -0.6660113871 | -0.8535014086 |
| 30 | C | 4.7621586781 | -1.4877111055 | -1.8189458784 |
| 31 | C | 3.2854262795 | -1.3812765205 | -1.9191900076 |
| 32 | C | 2.5678028776 | -0.4900275506 | -1.0801917529 |
| 33 | C | 3.2515254092 | 0.3278566438 | -0.1530256495 |
| 34 | C | 4.7296037003 | 0.2562216743 | -0.0178694306 |
| 35 | O | 5.4216974499 | -2.2452732765 | -2.5109689966 |
| 36 | O | 5.3530645094 | 0.9412485843 | 0.7778335736 |
| 37 | C | 1.1420384611 | -0.4185973272 | -1.1613057911 |
| 38 | C | 0.4346007230 | 0.4675083554 | -0.2985590738 |
| 39 | C | 1.1468820722 | 1.2712749241 | 0.5947980634 |
| 40 | C | 2.5549961935 | 1.1978191745 | 0.6648481162 |
| 41 | C | 2.6132762276 | -2.1766744476 | -2.8279087264 |
| 42 | C | 1.2074975816 | -2.1073226260 | -2.9201577615 |
| 43 | C | 0.4879956644 | -1.2558979757 | -2.1084739331 |
| 44 | C | -1.0577420273 | 0.5323103181 | -0.3188832827 |
| 45 | C | -1.7021744722 | 1.4455479032 | -1.1558118252 |
| 46 | C | -3.1110920305 | 1.5141474661 | -1.1944879114 |
| 47 | C | -3.8747327062 | 0.6798541851 | -0.3992732154 |
| 48 | C | -3.2619190434 | -0.2547623298 | 0.4678266233 |
| 49 | C | -1.8343579021 | -0.3293464538 | 0.5103923371 |
| 50 | C | -4.0392107839 | -1.1158772009 | 1.2890261010 |
| 51 | C | -3.4265053189 | -2.0243837762 | 2.1315388757 |
| 52 | C | -2.0197108700 | -2.1010120213 | 2.1798699893 |
| 53 | C | -1.2434490984 | -1.2781852863 | 1.3916400881 |
| 54 | C | -5.3488575157 | 0.7838084590 | -0.4860238454 |
| 55 | N | -6.0554585536 | -0.1030774820 | 0.3554174820 |
| 56 | C | -5.5166387885 | -1.0590521727 | 1.2520664176 |
| 57 | O | -6.2546813072 | -1.7596046036 | 1.9178411796 |
| 58 | O | -5.9433709589 | 1.5576339823 | -1.2131537428 |
| 59 | O | -0.8976794367 | 2.2348889404 | -1.9166128412 |
| 60 | C | -1.4936369051 | 3.2053563147 | -2.7701812683 |
| 61 | O | 0.4042046792 | 2.0951401476 | 1.3827776473 |
| 62 | C | 1.0685484472 | 2.9254224701 | 2.3287594685 |
| 63 | N | -7.4350343215 | -0.0418599949 | 0.2880648814 |
| 64 | C | -8.0503772911 | -0.8970684392 | -0.6539480970 |
| 65 | C | -9.5323645822 | -0.8271426577 | -0.6917617471 |
| 66 | C | -10.2295820994 | 0.0530770764 | 0.1735829007 |
| 67 | C | -9.5469350892 | 0.8893599445 | 1.0925203258 |
| 68 | C | -8.0652148510 | 0.8584508126 | 1.1767103480 |
| 69 | C | -10.2294286060 | -1.6319106921 | -1.5782343926 |
| 70 | C | -11.6348491093 | -1.5873563794 | -1.6290957796 |
| 71 | C | -12.3403044195 | -0.7372890410 | -0.7931168807 |
| 72 | C | -11.6500757973 | 0.0967495386 | 0.1196816061 |
| 73 | C | -12.3550609353 | 0.9719459099 | 0.9812601774 |
| 74 | C | -11.6637604089 | 1.7810248578 | 1.8681869488 |
| 75 | C | -10.2583679551 | 1.7392339855 | 1.9237758091 |
| 76 | C | -13.8274720935 | -0.7105048362 | -0.8575569077 |
| 77 | N | -14.4826830471 | 0.1715738678 | 0.0239695510 |
| 78 | C | -13.8425366912 | 1.0245799125 | 0.9442652295 |
| 79 | O | -7.4137781582 | 1.5423911054 | 1.9403863999 |
| 80 | O | -7.3863304395 | -1.6243780043 | -1.3651353746 |
| 81 | O | -14.4863080466 | 1.7636507284 | 1.6701717773 |
| 82 | O | -14.4587532940 | -1.4137301641 | -1.6285887437 |
| 83 | H | 13.2287803268 | -4.0608672686 | -0.2782598903 |
| 84 | H | 10.8147273477 | -3.5074012834 | -0.4817839433 |
| 85 | H | 10.6061877074 | 1.3215717718 | 0.1476655403 |
| 86 | H | 12.3200154972 | 3.0441649731 | 0.5444095140 |
| 87 | H | 17.7277436849 | -2.7783561345 | 0.3112943256 |
| 88 | H | 19.4450214390 | -1.0427783655 | 0.7201657738 |
| 89 | H | 20.0369729431 | 1.4350585457 | 1.1112450688 |
| 90 | H | 19.2481993232 | 3.7646734310 | 1.3338029655 |
| 91 | H | 16.8241839516 | 4.3267990303 | 1.1462052377 |
| 92 | H | 9.7608276840 | 0.2566750838 | -1.9837265800 |
| 93 | H | 7.3153459614 | 0.5635634847 | -2.2549578969 |
| 94 | H | 6.6032415652 | -2.2364365495 | 0.9131753330 |
| 95 | H | 9.0510635008 | -2.5316171771 | 1.2075068357 |
| 96 | H | 3.1244851560 | 1.8043788300 | 1.3583326663 |
| 97 | H | 3.1862713883 | -2.8497640098 | -3.4568973491 |
| 98 | H | 0.6877197155 | -2.7358659067 | -3.6374597085 |
| 99 | H | -0.5935485540 | -1.2175981218 | -2.1857811257 |
| 100 | H | -1.5453319798 | -2.8169111242 | 2.8444997194 |
| 101 | H | -4.0472344406 | -2.6691643958 | 2.7446700507 |
| 102 | H | -3.6299111919 | 2.2107281860 | -1.8415862918 |
| 103 | H | -9.6698752605 | -2.2959764226 | -2.2285579442 |
| 104 | H | -12.1876952227 | -2.2159803368 | -2.3190106865 |
| 105 | H | -12.2276333757 | 2.4423065219 | 2.5174649976 |
| 106 | H | -9.7100120229 | 2.3676272989 | 2.6176122581 |
| 107 | H | -0.6615671309 | 3.7180389509 | -3.2550116658 |
| 108 | H | -2.0871745086 | 3.9308496949 | -2.2005969918 |
| 109 | H | -2.1258632051 | 2.7360948454 | -3.5341004328 |
| 110 | H | 0.2776681038 | 3.4769638426 | 2.8391938646 |
| 111 | H | 1.7472158141 | 3.6325006437 | 1.8361662373 |
| 112 | H | 1.6297952489 | 2.3324710646 | 3.0614662987 |
| 113 | C | -15.9354808633 | 0.1958305600 | -0.0122703144 |
| 114 | C | -16.5758042831 | 1.1033784824 | -0.8669327383 |
| 115 | C | -15.7928075366 | 2.0518727685 | -1.7422349671 |
| 116 | C | -17.9746602124 | 1.1066726916 | -0.8850872666 |
| 117 | H | -18.4915017884 | 1.8025894620 | -1.5411198600 |
| 118 | C | -18.7030410777 | 0.2356850308 | -0.0782102109 |
| 119 | H | -19.7893125171 | 0.2512253056 | -0.1042471108 |
| 120 | C | -18.0395108006 | -0.6552508471 | 0.7620667992 |
| 121 | H | -18.6075101702 | -1.3352368184 | 1.3917472836 |
| 122 | C | -16.6420756964 | -0.6920872765 | 0.8108531924 |
| 123 | C | -15.9280367622 | -1.6609388888 | 1.7224049060 |
| 124 | H | -15.1387130339 | 1.5127210430 | -2.4381779245 |
| 125 | H | -15.1632655156 | 2.7231164956 | -1.1456869805 |
| 126 | H | -16.4699504255 | 2.6725081560 | -2.3358858926 |
| 127 | H | -15.2868386786 | -1.1393352003 | 2.4429976006 |
| 128 | H | -15.2954478879 | -2.3551657409 | 1.1561325625 |
| 129 | H | -16.6490947469 | -2.2564324861 | 2.2895347957 |
| 130 | H | -0.1618514242 | -1.3475515026 | 1.4361573513 |

**Table S4. Optimized cartesian coordinates** for (*S*)-**1**-$\boldsymbol{h_9}$.

**2-*h*$_9$** Coordinates (Angstroms)

| ATOM | | X | Y | Z |
|---|---|---|---|---|
| 1 | C | 16.2389988065 | -1.2085745481 | -0.2077704691 |
| 2 | C | 17.1878902191 | -0.2059808573 | -0.2793452660 |
| 3 | C | 16.8299273794 | 1.1742023226 | -0.3170548233 |
| 4 | C | 15.4410854736 | 1.4855056127 | -0.2778226179 |
| 5 | C | 14.4840152293 | 0.4491779158 | -0.2063419279 |
| 6 | C | 14.8694027171 | -0.8737099890 | -0.1706059345 |
| 7 | C | 17.7547747970 | 2.2498694870 | -0.3903938501 |
| 8 | C | 17.2985198747 | 3.5518039724 | -0.4212719286 |
| 9 | C | 15.9175781122 | 3.8614500529 | -0.3816585818 |
| 10 | C | 15.0010803418 | 2.8342174542 | -0.3105513955 |
| 11 | C | 13.0970326691 | 0.8124965246 | -0.1726387368 |
| 12 | C | 12.1384555570 | -0.2234624144 | -0.1015353339 |
| 13 | C | 12.5875519895 | -1.5696994087 | -0.0576249332 |
| 14 | O | 13.9335760099 | -1.8865876429 | -0.0994500212 |
| 15 | O | 13.6524933722 | 3.1509131251 | -0.2718134021 |
| 16 | C | 12.7171148087 | 2.1380387642 | -0.2036224147 |
| 17 | C | 10.7480550103 | 0.0871085227 | -0.0637994574 |
| 18 | C | 9.8083244508 | -0.9918796366 | 0.0109132387 |
| 19 | C | 10.2952627338 | -2.2888995449 | 0.0684673186 |
| 20 | C | 11.6738815602 | -2.5962097871 | 0.0326865914 |
| 21 | C | 11.3488373573 | 2.4716038626 | -0.1555441801 |
| 22 | C | 10.3974970273 | 1.4700278653 | -0.0848704088 |
| 23 | C | 8.3317252197 | -0.7595181719 | 0.0504412720 |
| 24 | C | 7.6012106376 | -0.3776154336 | -1.0944429734 |
| 25 | C | 6.2142493683 | -0.2425632951 | -0.9742137100 |
| 26 | C | 5.5296733140 | -0.4788391285 | 0.2250569668 |
| 27 | C | 6.2584542969 | -0.8476047203 | 1.3740350935 |
| 28 | C | 7.6454380811 | -0.9796441927 | 1.2522427729 |
| 29 | C | 4.0436434343 | -0.3120409793 | 0.2473807510 |
| 30 | N | -6.6617074487 | -0.0871117428 | 0.0317899129 |
| 31 | C | -7.1989954716 | 1.1549239433 | 0.4329140564 |
| 32 | C | -8.6824337269 | 1.2765791978 | 0.4028423851 |
| 33 | C | -9.4792319116 | 0.1863298400 | -0.0227609054 |
| 34 | C | -8.8909751873 | -1.0360746132 | -0.4270176042 |
| 35 | C | -7.4118089135 | -1.1949965597 | -0.4147551044 |
| 36 | C | -9.2815200505 | 2.4639669370 | 0.7924199155 |
| 37 | C | -10.6826414153 | 2.5973324631 | 0.7704126737 |
| 38 | C | -11.4830388262 | 1.5438361414 | 0.3577295753 |
| 39 | C | -10.8943243035 | 0.3209156682 | -0.0450258287 |
| 40 | C | -11.6904827031 | -0.7703307791 | -0.4691171980 |
| 41 | C | -11.0915596918 | -1.9579032258 | -0.8584565509 |
| 42 | C | -9.6905937706 | -2.0903281668 | -0.8389550103 |
| 43 | C | -12.9625439039 | 1.7029380639 | 0.3417135964 |
| 44 | N | -13.7111299304 | 0.5942052784 | -0.1030819200 |
| 45 | C | -13.1733751428 | -0.6498257526 | -0.4960251677 |
| 46 | O | -6.8637861581 | -2.2227158779 | -0.7751323587 |
| 47 | O | -6.4756912410 | 2.0703474716 | 0.7842692923 |
| 48 | O | -13.8970289769 | -1.5688623336 | -0.8389560481 |
| 49 | C | -15.1546589828 | 0.7392745158 | -0.1546807916 |
| 50 | O | -13.5120837213 | 2.7327688242 | 0.6951322293 |
| 51 | C | 3.4947464923 | 0.9611792279 | 0.2088086728 |
| 52 | C | 2.1019230158 | 1.1671133462 | 0.2190926194 |
| 53 | C | 1.2359920549 | 0.0904167814 | 0.2636186679 |
| 54 | C | 1.7530819681 | -1.2292276265 | 0.2908756845 |
| 55 | C | 3.1668059714 | -1.4460190393 | 0.2771093727 |
| 56 | C | 0.8712316853 | -2.3401449643 | 0.3248409869 |
| 57 | C | 1.3699699198 | -3.6304160318 | 0.3363252085 |
| 58 | C | 2.7607386249 | -3.8514261209 | 0.3063589431 |
| 59 | C | 3.6384023920 | -2.7862875641 | 0.2747457507 |
| 60 | C | -0.2275802471 | 0.3369229905 | 0.2716789316 |
| 61 | N | -1.0566731497 | -0.8064092500 | 0.2993107798 |
| 62 | C | -0.5994632762 | -2.1396610863 | 0.3439366068 |
| 63 | O | -1.3907526830 | -3.0683567764 | 0.3976357743 |
| 64 | O | -0.7106143978 | 1.4572431884 | 0.2538963635 |
| 65 | C | -2.4898471990 | -0.5990471072 | 0.2617886624 |
| 66 | C | -3.2513804196 | -0.7151961922 | 1.4315379359 |
| 67 | C | -4.6305922052 | -0.5316386760 | 1.3077274606 |
| 68 | C | -5.2197029881 | -0.2407047259 | 0.0810921212 |
| 69 | C | -4.4574502508 | -0.1041530699 | -1.0853433086 |
| 70 | C | -3.0795013495 | -0.2952408689 | -0.9620516290 |
| 71 | C | -15.9194129176 | 0.5629222038 | 1.0044927766 |
| 72 | C | -17.3051238910 | 0.7184582179 | 0.8756971520 |
| 73 | C | -17.8968008267 | 1.0286485001 | -0.3460874637 |
| 74 | C | -17.1231004237 | 1.1976762310 | -1.5010089538 |
| 75 | C | -15.7394902015 | 1.0490573728 | -1.3807963998 |
| 76 | C | 5.5898402420 | -1.0765189868 | 2.7111253811 |
| 77 | C | 8.2689661013 | -0.1523716750 | -2.4318897687 |
| 78 | C | -15.2913606761 | 0.2318866574 | 2.3353903073 |
| 79 | C | -17.7635765474 | 1.5187317461 | -2.8311611111 |
| 80 | C | -5.0808412467 | 0.2132739417 | -2.4212502578 |
| 81 | C | -2.6294808604 | -1.0459030744 | 2.7643896152 |
| 82 | H | 16.5283274190 | -2.2543837020 | -0.1794858149 |
| 83 | H | 18.2411927950 | -0.4720913366 | -0.3079015850 |
| 84 | H | 18.8198435370 | 2.0373044837 | -0.4218104211 |
| 85 | H | 18.0112582311 | 4.3699210677 | -0.4771319913 |
| 86 | H | 15.5718029689 | 4.8895139374 | -0.4063840345 |
| 87 | H | 12.0188211331 | -3.6240983751 | 0.0710325439 |
| 88 | H | 9.5852615723 | -3.1089357649 | 0.1317488983 |
| 89 | H | 9.3475263123 | 1.7404292283 | -0.0328548659 |
| 90 | H | 11.0587329631 | 3.5175765206 | -0.1685301742 |
| 91 | H | 8.2228109938 | -1.2577071673 | 2.1308429273 |
| 92 | H | 5.6374337608 | 0.0341247450 | -1.8537879620 |
| 93 | H | 4.1597060550 | 1.8194832807 | 0.1795617311 |
| 94 | H | 1.6856988906 | 2.1689789140 | 0.1948086659 |
| 95 | H | 3.1420873980 | -4.8683904838 | 0.3037284137 |
| 96 | H | 0.6681482981 | -4.4574409110 | 0.3619468023 |
| 97 | H | -2.4475344558 | -0.2074632570 | -1.8408175044 |
| 98 | H | -5.2627547834 | -0.6206601743 | 2.1864888851 |
| 99 | H | -9.2133519982 | -3.0156829381 | -1.1435479711 |
| 100 | H | -11.7247431077 | -2.7788973905 | -1.1776567537 |
| 101 | H | -8.6483949250 | 3.2843249520 | 1.1134612214 |
| 102 | H | -11.1590379521 | 3.5233938711 | 1.0741983013 |
| 103 | H | -15.1008663835 | 1.1732882006 | -2.2512005937 |
| 104 | H | -18.9775269834 | 1.1395286334 | -0.4033115290 |
| 105 | H | -17.9302390061 | 0.5919175915 | 1.7563413867 |
| 106 | H | 4.7088835600 | -2.9620638588 | 0.2414674872 |
| 107 | H | 4.9571173014 | -0.2278118206 | 2.9955067481 |
| 108 | H | 6.3367414518 | -1.2182586100 | 3.4979741818 |
| 109 | H | 4.9445162717 | -1.9634368845 | 2.7024520535 |
| 110 | H | 8.8909949801 | -1.0087279264 | -2.7162848522 |
| 111 | H | 8.9269716201 | 0.7248979684 | -2.4193180398 |
| 112 | H | 7.5234146608 | 0.0018706665 | -3.2179373251 |
| 113 | H | -5.7028355985 | 1.1146625473 | -2.3728422750 |
| 114 | H | -5.7136727762 | -0.6124845956 | -2.7675856034 |
| 115 | H | -4.3083483820 | 0.3824725330 | -3.1768905609 |
| 116 | H | -2.2492885833 | -2.0740594439 | 2.7707459835 |
| 117 | H | -3.3636218934 | -0.9506510898 | 3.5697020181 |
| 118 | H | -1.7903299813 | -0.3791517595 | 2.9937817489 |
| 119 | H | -14.6693244671 | -0.6692519090 | 2.2773927887 |
| 120 | H | -16.0620546533 | 0.0555260727 | 3.0915518177 |
| 121 | H | -14.6568904844 | 1.0529112394 | 2.6901940205 |
| 122 | H | -17.0101458972 | 1.7247154823 | -3.5977846523 |
| 123 | H | -18.4179315116 | 2.3957306539 | -2.7593981379 |
| 124 | H | -18.3799122012 | 0.6840476547 | -3.1886897819 |

**Table S5. Optimized cartesian coordinates** for **2-*$h_9$***.

| $\boldsymbol{\theta}$ | $\boldsymbol{I_{CISS}}$ | $\boldsymbol{I_S}$ | $\boldsymbol{I_{13}}$ |
|---|---|---|---|
| 0 | $\frac{1}{4}\cos^2\phi$ | $\frac{1}{2}\sin^2\phi\cos^2\phi$ | $\frac{1}{4}\cos^2\phi\,(1-\cos 2\phi\cos 2\chi)$ |
| 90 | $-\frac{1}{8}\cos 2\phi\cos^2\phi$ | $\frac{1}{2}\sin^2\phi\cos^2\phi$ | $\frac{1}{4}\cos^2\phi\,(\cos 2\chi-\cos 2\phi\cos^2\chi)$ |

**Table S6. Summary of the different intensities for transition 1↔3** assuming 100% CISS, pure singlet and partial CISS efficiency (see Eqs. S16, S15, S13).

| | |
|---|---|
| $g_{NDI}$ | $(2.0034,\ 2.0041,\ 2.0044)$ |
| $g_{PXX}$ | $(2.0031,\ 2.0044,\ 2.0046)$ |
| $r_{DA}$ | 2.5 nm |
| $\sigma_D$ | 0.35 nm |
| $a_{NDI}$ | 6.3 MHz |
| $a_{PXX}$ | 10 MHz |
| $\Delta B_0$ | 0.4 mT |

**Table S7. Summary of the parameters used for the simulations of Fig. 4**, where $\sigma_D$ is the width of the gaussian distribution of DA distances (used to compute $D(\theta)$, Eq. S2) and $\Delta B_0$ is the additional gaussian broadening used to convolute the spectra and accounting for inhomogeneous broadening of the remaining parameters. Parameters are given in the laboratory frame, with the magnetic field along z.